\documentclass[english]{article}
\usepackage[T1]{fontenc}
\usepackage[latin9]{inputenc}
\pdfoutput=1
\usepackage{geometry}
\usepackage{color}
\usepackage{float}
\usepackage{calc}
\usepackage{textcomp}
\usepackage{mathrsfs}
\usepackage{amsmath}
\usepackage{amsthm}
\usepackage{amssymb}
\usepackage{graphicx}

\makeatletter

\providecommand{\tabularnewline}{\\}

\theoremstyle{plain}
\newtheorem{thm}{\protect\theoremname}
\theoremstyle{remark}
\newtheorem{rem}[thm]{\protect\remarkname}

\usepackage{ dsfont }
\usepackage[all]{xy}
\usepackage{cite}
\usepackage{colonequals}

\usepackage{textgreek}

\usepackage{tikz}

\makeatletter
\let\@fnsymbol\@arabic
\makeatother

\RequirePackage{mathtools}
\global\long\def\defi{\vcentcolon =}

\AtBeginDocument{

}

\makeatother

\usepackage{babel}
\providecommand{\remarkname}{Remark}
\providecommand{\theoremname}{Theorem}

\begin{document}
\global\long\def\u{u}
\global\long\def\p{P}
\global\long\def\X{X}
\global\long\def\me{\mu_{e}}
\global\long\def\sym{\textrm{sym}}
\global\long\def\grad{\nabla}
\global\long\def\le{\lambda_{e}}
\global\long\def\tr{\textrm{tr}}
\global\long\def\mc{\mu_{c}}
\global\long\def\skew{\textrm{skew}}
\global\long\def\curl{\textrm{Curl}}
\global\long\def\ac{\alpha_{c}}
\global\long\def\B{\mathscr{B}}
\global\long\def\R{\mathbb{R}}
\global\long\def\fr{\rightarrow}
\global\long\def\Q{\mathcal{Q}}
\global\long\def\A{\mathscr{A}}
\global\long\def\L{\mathscr{L}}
\global\long\def\D{\mathscr{D}}
\global\long\def\id{\mathds{1}}
\global\long\def\ds{\textrm{dev sym}}
\global\long\def\sph{\textrm{sph}}
\global\long\def\eg{\boldsymbol{\varepsilon}}
\global\long\def\aa{\boldsymbol{\alpha}}
\global\long\def\o{\boldsymbol{\omega}}
\global\long\def\ege{\boldsymbol{\varepsilon}_{e}}
\global\long\def\egp{\boldsymbol{\varepsilon}_{p}}
\global\long\def\punto{\,.\,}
\global\long\def\so{\mathfrak{so}\left(3\right)}
\global\long\def\Sym{\textrm{Sym}\left(3\right)}
\global\long\def\MM{\boldsymbol{\mathfrak{M}}}
\global\long\def\C{\mathbb{C}}
\global\long\def\gl{\mathfrak{gl}\left(3\right)}
\global\long\def\P{P}
\global\long\def\Lin{\textrm{Lin}}
\global\long\def\D{\boldsymbol{D}}
\global\long\def\a{\alpha}
\global\long\def\b{\beta}
\global\long\def\lle{\lambda_{e}}
\global\long\def\ce{\mathbb{C}_{e}}
\global\long\def\cm{\mathbb{C}_{\textrm{micro}}}
\global\long\def\cc{\mathbb{C}_{c}}
\global\long\def\axl{\textrm{axl}}
\global\long\def\dev{\textrm{dev}}
\global\long\def\Ls{\widehat{\mathbb{L}}}
\global\long\def\mh{\mu_{\textrm{micro}}}
\global\long\def\lh{\lambda_{\textrm{micro}}}
\global\long\def\vau{\omega_{1}^{int}}
\global\long\def\vad{\omega_{2}^{int}}
\global\long\def\axl{\textrm{axl}}
\global\long\def\Le{\mathbb{L}_{e}}
\global\long\def\Lc{\mathbb{L}_{c}}
\global\long\def\V{\mathbb{V}}
\global\long\def\as{\dagger}
\global\long\def\dd{\:\overset{{\scriptscriptstyle \triangledown}}{{\scriptscriptstyle \vartriangle}}\,}
\global\long\def\cM{\mathbb{C}_{\textrm{macro}}}
\global\long\def\mum{\mu_{\textrm{macro}}}
\global\long\def\lam{\lambda_{\textrm{macro}}}
\global\long\def\gi{\mathfrak{i}}

\title{\textbf{Effective description of anisotropic wave dispersion in mechanical
band-gap metamaterials via the relaxed micromorphic model}}

\author{Marco Valerio d\textquoteright Agostino\thanks{Marco Valerio d'Agostino, corresponding author, marco-valerio.dagostino@insa-lyon.fr,
GEOMAS, INSA-Lyon, Université de Lyon, 20 avenue Albert Einstein,
69621, Villeurbanne cedex, France}, \, Gabriele Barbagallo\thanks{Gabriele Barbagallo, gabriele.barbagallo@insa-lyon.fr, GEOMAS, INSA-Lyon, Universitité de Lyon, 20 avenue Albert Einstein,
69621, Villeurbanne cedex, France}, \, Ionel-Dumitrel Ghiba\thanks{Ionel-Dumitrel Ghiba, dumitrel.ghiba@uaic.ro, Alexandru Ioan Cuza
University of Ia\c{s}i, Department of Mathematics, Blvd. Carol I,
no. 11, 700506 Ia\c{s}i, Romania; and Octav Mayer Institute of Mathematics
of the Romanian Academy, Ia\c{s}i Branch, 700505 Ia\c{s}i.}, \, Bernhard Eidel\thanks{Bernhard Eidel, bernhard.eidel@uni-siegen.de, Universität Siegen,
Institut für Mechanik, Heisenberg-group, Paul-Bonatz-Straße 9-11 57076
Siegen, Germany} ,\\
\, Patrizio Neff\,\thanks{Patrizio Neff, patrizio.neff@uni-due.de, Head of Chair for Nonlinear
Analysis and Modelling, Fakultät für Mathematik, Universität Duisburg-Essen,
Mathematik-Carrée, Thea-Leymann-Straße 9, 45127 Essen, Germany}$\;$$\;$and Angela Madeo\thanks{Angela Madeo, angela.madeo@insa-lyon.fr, GEOMAS, INSA-Lyon, Université
de Lyon, 20 avenue Albert Einstein, 69621, Villeurbanne cedex, France}}
\maketitle
\begin{abstract}
In this paper the relaxed micromorphic material model for anisotropic
elasticity is used to describe the dynamical behavior of a band-gap
metamaterial with tetragonal symmetry. Unlike other continuum models
(Cauchy, Cosserat, second gradient, classical Mindlin-Eringen micromorphic
etc.), the relaxed micromorphic model is endowed to capture the
main microscopic and macroscopic characteristics of the targeted metamaterial,
namely, stiffness, anisotropy, dispersion and band-gaps.

The simple structure of our material model, which simultaneously lives
on a micro-, a meso- and a macroscopic scale, requires only the identification
of a limited number of frequency-independent and thus truly constitutive
parameters, valid for both static and wave-propagation analyses in
the plane. The static macro- and micro-parameters are identified by
numerical homogenization in static tests on the unit-cell level in \cite{d2019Identification}. 

The remaining inertia parameters for dynamical analyses are calibrated
on the dispersion curves of the same metamaterial as obtained by a
classical Bloch-Floquet analysis for two wave directions. 

We demonstrate via polar plots that the obtained material parameters
describe very well the response of the structural material for all
wave directions in the plane, thus covering the complete panorama
of anisotropy of the targeted metamaterial.
\end{abstract}
\addtocounter{footnote}{5} \vspace{6mm}

\vspace{2mm}
\textbf{}

\newpage{}

\tableofcontents{}

\section{Introduction}

\addtocounter{footnote}{-5} 

\subsection{Describing metamaterials as generalized continua }

Engineering metamaterials showing exotic behaviors with respect to
both mechanical and electromagnetic wave propagation have recently been attracting growing attention for their numerous possible astonishing
applications \cite{armenise2010phononic,steurer2007photonic,findeisen2017characteristics,lee2011negative}.
Actually, materials which are able to \textquotedblleft stop\textquotedblright{}
or \textquotedblleft bend\textquotedblright{} the propagation of waves
of light or sound with no energetic cost could suddenly disclose rapid
and unimaginable technological advancements. Metamaterials exhibiting
such unorthodox behaviors are obtained by suitably assembling different
microstructural components in such a way that the resulting macroscopic
material possesses completely new properties with respect to the original
one. For example, in \cite{dong2017topology} a topology optimization procedure is used to design 2D single-phase, anisotropic elastic metamaterials with broadband double-negative effective material properties which exhibit a superlensing effect at the deep-subwavelength scale.

By their intrinsic nature, metamaterials show strong heterogeneities
at the level of the microstructure and, except for few particular
cases, their mechanical behavior is definitely anisotropic. Depending
on their degree of anisotropy, band-gap metamaterials can exhibit
one or both of the following behaviors: (i) anisotropic behavior with respect to deformation (the deformation
patterns vary when varying the direction of application of the externally
applied loads), (ii) anisotropic behavior with respect to band-gap properties (the width
of the band-gap varies when varying the direction of propagation of
the travelling wave).

Thus, the description of anisotropy in metamaterials is a challenging
issue, given that extra innovative applications could be conceived.
In fact, a metamaterial in which different modes propagate with different
speeds when changing the direction of propagation could be fruitfully
employed as wave-guides or wave filters. 

The need of a homogenized model which is able to account for anisotropy
in band-gap metamaterials at large scale is of great concern for the
engineering scientific community. Indeed, the ultimate task of an
engineer is that of dealing with models which are able to describe
the overall macroscopic behavior of (meta-)materials in the most simplified
possible way in order to proceed towards the conception of morphologically
complex engineering (meta-)structures.
In this spirit, we could find several approaches in literature as that proposed in \cite{charlotte2012lattice}.

As Green abandoned any attempt to relate the elastic behavior of materials
to their molecular arrangements, we abandon any effort to connect
``a priori'' the elastic behavior of metamaterials to the arrangement
of their constituting elements.
Nevertheless, when the best macroscopic model for the description
of the mechanical behavior of metamaterials is selected it will
be easy to connect ``a posteriori'' some of its elastic parameters
to the specific properties of the unit-cell. Hence our primary goal
is that of establishing which continuum model has to be used to describe
the mechanical behavior of (isotropic and anisotropic) metamaterials
at the macroscopic scale.

To this aim, we want to start from the easiest possible ``experimental''
evidence and then try to build macroscopic strain and kinetic energy
densities which are able to account for the phenomena we are interested
in.

Our macroscopic primary observation is the typical behavior of the
dispersion curves in a metamaterial (see, e.g., Fig.\,\ref{fig:Comsol model}).
Since the considered metamaterial is not isotropic, such dispersion
curves vary when changing the direction of propagation of the travelling
wave (Fig.\,\ref{fig:Comsol model}(a) and (b)).

We start from the observation that the typical dispersion curves of
a given metamaterial show different branches that can be classified
roughly as follows: 1) acoustic branches (starting from the origin) which, very close to
the origin, are well approximated by straight lines that coincide
with the straight lines obtained by classical linear elasticity. Such
branches, at least for small wavenumbers (large wavelengths) are related
to the macroscopic modes of vibration of the unit-cell, 2) optic branches (starting from cut-off values of the frequency) which
are related to the modes of vibration of the microstructure inside
the unit-cell.

We then proceed trying to find the simplest possible continuum model
which allows us to account for the behavior of all such dispersion
curves. It is clear that classical elasticity is too restrictive to
accomplish this task. Indeed, in the fully anisotropic case, classical
elasticity features at most 3 different acoustic dispersion curves
which are straight lines the slopes of which gives the speed of propagation
of compression and shear waves inside the material. In other words,
in classical elastic solids waves propagate with the same speed for
any wavelength: a classical elastic solid is said to be ``non-dispersive''.
To proceed in the right direction and find a good candidate for the
``best'' continuum model for metamaterials, we need to introduce
in the model two fundamental things: 1)the ability of describing dispersive behaviors (the acoustic curves
are not straight lines but curves), 2) the ability of introducing extra ``optic'' curves related to the
vibration of the microstructure (often these curves show dispersive
behaviors).
As we anticipated, none of these two features can be obtained by classical
elasticity, but the answer must be found in the realm of so-called
``enriched'' continuum models. Nevertheless, the choice of the ``best''
enriched continuum model is not a trivial task, considering that a
huge variety of such models is present in the literature.

A way to approach point 1.\,(i.e., introducing dispersive behavior
for acoustic modes in the picture) could be that of using so-called
higher-gradient theories. Indeed, it is known that considering a strain
energy density which depends not only on $\varepsilon=\sym\,\nabla u$,
but also on its gradient $\nabla\varepsilon$ allows for obtaining
governing equations of higher order than those of classical elasticity.
This, in terms of dispersion curves, means that the acoustic lines
are not straight, but can show some dispersion (see, e.g., \cite{placidi2014reflection,dellisola2012linear,Rosi2017validity}).

Moreover, generalizing the constitutive form of the strain energy
density to account for anisotropic behaviors (see for example \cite{smyshlyaev2009propagation}) not only on the first
gradient, but also on the second gradient terms (see for example \cite{smyshlyaev2000rigorous} for a rigorous derivation of strain gradient effects in periodic media), qualitative anisotropic
patterns for the wave speeds associated to the acoustic compression
and shear waves can be obtained (see \cite{rosi2016anisotropic}).

Approaching the modeling of metamaterials through second (or higher)
order theories has, at least, two limitations, namely: i) no optic branches can be described, but only some dispersion in the
acoustic curves, ii) the treatment of anisotropy in the framework of higher gradient continua
quickly becomes uselessly complicated. Indeed, the need of introducing
non-classical elastic tensors (of the sixth order against the fourth
order of classical elasticity) arises and the study of the class of
symmetries of such tensors introduces non-trivial technical difficulties.
On the other hand, the introduced complexity is not justified by a
true advantage in terms of enhanced description of the physical phenomena
concerning metamaterials: the only improvements with respect to classical
elasticity are the description of the dispersion for acoustic curves
and the description of anisotropy only for the first two (acoustic)
modes.

We will show in the remainder of this paper that both such informations,
as well as many extra features such as the description of optic modes,
anisotropy (also at higher frequencies) and band-gaps, can be obtained
in a much more simple fashion which does not need to invoke any new
theoretical framework with respect to the classical treatment of classes
of symmetries for classical elasticity.

\subsection{Isotropic modelling of metamaterials via the relaxed micromorphic model}

Having shown that second gradient models are not the right way to
answer to the need of an optimal enriched continuum material model
for metamaterials, the attention has to be shifted on so-called micromorphic
models. Micromorphic models feature an enriched kinematics with respect
to classical elasticity in the sense that extra degrees of freedom
are added to the continuum. The enriched kinematics thus consists
of the total macroscopic displacement vector $u$ plus a second order tensor
(generally not symmetric) $P$ which is known as micro-distortion
tensor. The simple fact of enriching the kinematics allows for the
possibility of describing extra (optic) dispersion curves, and thus
for including the effect of microstructure on the dynamical behavior
of heterogeneous materials (see, e.g., \cite{chen2003connecting,chen2003determining}).
The properties and the shape of such curves then depend on the constitutive
choice that one makes for the strain energy and kinetic energy densities.
The true difficulty is thus that of making a ``smart'' selection
of such constitutive choices so that:
\begin{itemize}
\item the expressions of both the strain energy and the kinetic energy densities
are the easiest possible, avoiding any unuseful complexification,
\item such expressions still allow to describe the macroscopic phenomena
we are interested in (dispersion and anisotropy for acoustic and optic
modes, band-gaps,...).
\end{itemize}
We have already addressed the problem of selecting the possible model
for description of metamaterials' elasticity for the isotropic case
(see \cite{madeo2016reflection,madeo2016first,madeo2016complete,madeo2015wave,dagostino2016panorama,aivaliotis2018low,aivaliotis2019microstructure}).
The answer we found is that this optimal choice is given by the so-called
\textbf{relaxed micromorphic model}.
We showed that this model has the following advantages:
\begin{itemize}
\item it produces the smallest possible number of elastic parameters in
the strain energy density with respect to classical isotropic elasticity, 
\item all the introduced homogenized parameters are true material constants
(exactly as the averaged Young's modulus and Poisson's ratio) since
they do not depend on frequency, as it is instead the case for classical
dynamic homogenization results (see, e.g., \cite{antonakakis2013high,nemat2011homogenization,nemat2011overall}),
\item the splitting of the displacement gradient $\nabla u$ and the micro-distortion
$P$ in their othogonal $\sym$ and $\skew$ part allows from one side, to naturally
extend classical elasticity and, from the other side, to isolate macro
and micro deformation modes related to distortions and rotations,
respectively,
\item it allows the description of dispersion (wave speed varying with the
considered wavelength) not only for the acoustic modes, but also for
the optic modes at higher frequencies,
\item it allows, when desired, the description of non-localities in metamaterials
thanks to the term $\curl\,P$ which includes combinations of space
derivatives of the micro-distortion $P$. In this respect, we have
to remember that such a constitutive choice is much less restrictive
compared to classical Mindlin-Eringen micromorphic models featuring
the whole gradient $\nabla P$ of $P$ in the strain energy density
\cite{eringen1999microcontinuum,mindlin1964micro}. We indeed showed
(see \cite{neff2014unifying,neff2016real,madeo2016reflection,madeo2016first,madeo2016role,madeo2017modeling,madeo2016complete,madeo2015wave,madeo2014band})
that the non-localities introduced by $\nabla P$ are so strong that
sometimes they preclude the micromorphic model from describing essential
features such as band-gaps. We found the relaxed micromorphic model
to be the best compromise between the description of non-localities
and the possibility of allowing realistic band-gap behaviors.
\end{itemize}

\subsection{Anisotropic modelling approach in this work}

At this point, the remaining question is ``how to select
the optimal model for the description of anisotropy in metamaterials''?

We started answering this question in \cite{barbagallo2016transparent}
where the generalization of the relaxed micromorphic model to the
anisotropic case was presented. One of the main advantages of this
model, as we will see in the remainder of this paper, is that of describing
the anisotropy related to the microstructure. To this aim, no cumbersome
treatment related to the classes of symmetries of higher order tensors
is needed since, in our model, everything can be recast in the classical
study of the classes of symmetry of the classical fourth order elasticity
tensor.

We end up with the simplest possible continuum model which is able
to describe simultaneously macro and micro anisotropies in metamaterials, the dispersion and band-gaps and the non-local effects.

We prove the efficacy of this relatively simplified model by superimposing
the dispersion curves of our model with the ``phenomenological evidence''
which in this paper we suppose to be the dispersion curves of a given
anisotropic (tetragonal) metamaterial obtained by so-called Bloch-Floquet
analysis \cite{floquet1883equations,bloch1929quantenmechanik}.

We will also show that our model is able to recover the behavior of
the phase velocity as a function of the direction of propagation of
the travelling wave not only for the first two acoustic modes, but
also for the optic modes.

We find an excellent agreement with the phenomenological evidence,
often not only for large wavelengths, but also for wavelengths which
become relatively close to the size of the unit-cell.

In the present paper, while developing a relaxed micromorphic theoretical
framework which is capable to generally treat full anisotropy in metamaterials
(also based on the results obtained in \cite{barbagallo2016transparent}),
we will present a first application to an actual metamaterial with
a low degree of anisotropy (for other applications see \cite{barbagallo2018relaxed,aivaliotis2019broadband}). More specifically, we select a metamaterial
with a particular microstructure for which the band-gap is almost
isotropic (not varying with the direction of propagation of the traveling
wave), but which has an anisotropic (tetragonal) elastic behavior.
As a second step, we use the anisotropic relaxed micromorphic model
to reproduce both the patterns of the dispersion curves and of the
phase velocity as function of the angle giving the direction of propagation
of the travelling wave. 

We compare the obtained results with the analogous ones issued by
a classical Bloch wave analysis of the same metamaterial, showing
that the relaxed micromorphic model is able to catch the main features
of the mechanical behavior of such metamaterial, namely: the overall patterns of the dispersion curves as function of the direction
of propagation, the polar plots of the phase velocity and the band-gap characteristics. 

The proposed approach has been used in other papers to study interesting mechanical problems involving anisotropic microstructured materials as in \cite{barbagallo2018relaxed}.

\medskip

\noindent This paper is now structured as follows
\begin{itemize}
\item in Section 2 we introduce the notation used in the paper, 
\item in Section 3 we present the general anisotropic relaxed micromorphic
model in a variational format and the governing system of PDEs,
\item in Section 4 we introduce the plane wave ansatz on the unknown kinematical
fields in order to show how it is possible to reduce the system of
governing PDEs to an algebraic problem, describing the procedures
to derive the dispersion curves,
\item in Section 5 we recall the basics of the Bloch-Floquet theory which allows to determine the wave propagation in a periodic medium by solving a spectral problem on the unit-cell,
\item in Section 6 we consider a particular periodic microstructure which
has tetragonal symmetry and we perform the Bloch-Floquet analysis
in order to derive the dispersion curves associated to the equivalent
continuum. The dispersion curves obtained with this method will be
used in following Sections to calibrate the parameters of the relaxed
micromorphic model,
\item in Section 7 we specialize the general framework of the anisotropic
relaxed micromorphic model presented in Section 3 to the particular
case of the tetragonal material symmetry,
\item in Section 8 we present in detail the fitting procedure that we used
to obtain the remaining parameters of the relaxed micromorphic model
which have not been determined by static arguments in \cite{d2019Identification}. This will be done
via the superposition of the dispersion curves obtained from our model
to those obtained via Bloch-Floquet analysis,
\item in Section 9 we show how the proposed anisotropic relaxed model is
able to catch the anisotropic behavior of the considered tetragonal
metamaterial. This is done by comparing the polar plots of the phase
velocity as obtained with the relaxed micromorphic model to those
obtained via Bloch-Floquet analysis. We show that a very good agreement
exists for all directions of propagation and for wavelength which
can become very small, even comparable to the size of the unit-cell.
\end{itemize}

\section{Notation}

Throughout this paper the Einstein convention of summation over repeated
indexes is used unless stated otherwise. We denote by $\R^{3\times3}$
the set of real $3\times3$ second order tensors and by $\R^{3\times3\times3}$
the set of real $3\times3\times3$ third order tensors. The standard
Euclidean scalar product on $\R^{3\times3}$ is given by $\left\langle X,Y\right\rangle {}_{\R^{3\times3}}=\tr(X\cdot Y^{T})$
and, thus, the Frobenius tensor norm is $\|X\|^{2}=\left\langle X,X\right\rangle {}_{\R^{3\times3}}$.
Moreover, the identity tensor on $\R^{3\times3}$ will be denoted
by $\mathds{1}$, so that $\tr(X)=\left\langle X,\mathds{1}\right\rangle $.
We adopt the usual abbreviations of Lie-algebra theory, i.e., $\Sym\colonequals\{X\in\R^{3\times3}\;|X^{T}=X\}$ denotes the vector-space
of all symmetric $3\times3$ matrices, $\so\colonequals\{X\in\R^{3\times3}\;|X^{T}=-X\}$ is the Lie-algebra
of skew symmetric tensors, $\mathfrak{sl}(3)\colonequals\{X\in\R^{3\times3}\;|\tr(X)=0\}$ is
the Lie-algebra of traceless tensors,
 $\R^{3\times3}\simeq\mathfrak{gl}(3)=\{\mathfrak{sl}(3)\cap\Sym\}\oplus\so\oplus\R\!\cdot\!\mathds{1}$
is the \emph{orthogonal Cartan-decomposition of the Lie-algebra $\mathfrak{gl}(3)$}.
In other words, for all $X\in\R^{3\times3}$, we consider the orthogonal
decomposition $X=\ds X+\skew X+1/3\,\mathrm{tr}(X)\,\mathds{1},\label{eq:cartan lie}$
where: $\sym\,X=\frac{1}{2}(X^{T}+X)\in\Sym$ is the symmetric part of $X$, $\skew\,X=\frac{1}{2}(X-X^{T})\in\so$ is the skew-symmetric part
of $X$, $\dev\,X=X-\frac{1}{3}\tr(X)\,\mathds{1}\in\mathfrak{sl}(3)$ is the
deviatoric (trace-free) part of $X$. Throughout all the paper we indicate: 
\begin{itemize}
\item without superscripts, i.e., $\,\C$, a classical fourth order tensor
acting only on symmetric matrices \\
 $\left(\C:\Sym\rightarrow\Sym\right)$ or skew-symmetric ones $\left(\cc:\so\rightarrow\so\right)$
,
\item with tilde, i.e., $\widetilde{\mathbb{C}}_{c}$, a second order tensor
$\widetilde{\mathbb{C}}_{c}:\R^{3}\rightarrow\R^{3}$ appearing as
elastic stiffness of certain coupling terms. 
\end{itemize}
The operation of simple contraction between tensors of suitable order
is denoted by $\cdot\;$; for example, 
\begin{align}
\left(\widetilde{\C}\cdot v\right)_{i}=\widetilde{\C}_{ij}v_{j}\,,\qquad\left(\widetilde{\C}\cdot X\right)_{ij}=\widetilde{\C}_{ih}X_{hj}\,.
\end{align}

\noindent Typical conventions for differential operations are used,
such as a comma followed by a subscript to denote the partial derivative
with respect to the corresponding Cartesian coordinate, i.e., \,$\left(\cdot\right)_{,j}=\frac{\partial(\cdot)}{\partial x_{j}}$.

\noindent The $\textrm{curl}$ of a vector field $v$ is defined as
$\left(\textrm{curl}\,v\right)_{i}=\varepsilon_{ijk}\,v_{k,j}$,
where $\varepsilon_{ijk}$ is the Levi-Civita third order permutation
tensor. Let $X\in\R^{3\times3}$ be a second order tensor field and
$X_{1},X_{2},X_{3}\in\R^{3}$ three vector fields such that
$X=\left( X_{1}^{T},X_{2}^{T},X_{3}^{T}\right)$. The $\curl$ of a smooth matrix field $X$ is defined as follows:
$\curl\,X=\big(\left(\textrm{curl}\,X_{1}\right)^{T},\left(\textrm{curl}\,X_{2}\right)^{T},\left(\textrm{curl}\,X_{3}\right)^{T}\big)^T $
or in index notation, $\left(\curl\,X\right)_{ij}=\varepsilon_{jmn}\,X_{in,m}$. For the iterated $\curl$ we have $\left(\curl\:\curl\:P\right)_{ij} =P_{im,jm}-P_{ij,mm}$.
The divergence $\textrm{div}\,v$ of a smooth vector field $v$ is
defined as $\textrm{div}\,v=v_{i,i}$ and the divergence $\textrm{Div}\,X$
of a tensor field $X\in\R^{3\times3}$ as
$\textrm{Div}\,X=\left( \textrm{div}\,X_{1},\textrm{div}\,X_{2},\textrm{div}\,X_{3}\right)^T$
or, in index notation, $\left(\textrm{Div}\,X\right)_{i}=X_{ij,j}$.

\section{Variational formulation of the relaxed micromorphic model}

\subsection{Constitutive assumptions on the Lagrangian and equations of motion}

For the Lagrangian energy density we assume the standard split into
kinetic minus potential energy:

\begin{equation}
\mathscr{L}\left(\u_{,t},\nabla u_{,t},\P_{,t},\nabla\u,\P,\curl\,\p\right)=J\left(\u_{,t},\nabla u_{,t},\P_{,t}\right)-W\left(\nabla\u,\P,\curl\,\p\right).
\end{equation}
When considering anisotropic linear elastic micromorphic media, as
reported in \cite{barbagallo2016transparent,neff2014unifying}, the
kinetic energy density and the potential one may take on the form
\begin{align}
J\left(\u_{,t},\nabla u_{,t},\P_{,t}\right) & =\frac{1}{2}\left\langle \rho\,u_{,t},u_{,t}\right\rangle +\frac{1}{2}\left\langle \mathbb{J}_{\textrm{micro}}\,\sym\,\P_{,t},\sym\,\P_{,t}\right\rangle +\frac{1}{2}\left\langle \mathbb{J}_{c}\,\skew\,\P_{,t},\skew\,\P_{,t}\right\rangle \label{kin energy}\\
 & \qquad+\frac{1}{2}\left\langle \mathbb{T}\,\sym\,\nabla u_{,t},\sym\,\nabla u_{,t}\right\rangle +\frac{1}{2}\left\langle \mathbb{T}_{c}\,\skew\,\nabla u_{,t},\skew\,\nabla u_{,t}\right\rangle ,\nonumber \\
W\left(\nabla\u,\P,\curl\,\p\right) & =\underbrace{\frac{1}{2}\left\langle \mathbb{C}_{e}\,\sym\left(\grad\u-P\right),\sym\left(\grad\u-P\right)\right\rangle _{\R^{3\times3}}}_{\textrm{anisotropic elastic - energy}}+\underbrace{\frac{1}{2}\left\langle \mathbb{C}_{\textrm{micro}}\,\sym\,P,\sym\,P\right\rangle _{\R^{3\times3}}}_{\textrm{micro - self - energy}}\label{pot energy}\\
 & \qquad+\underbrace{\frac{1}{2}\left\langle \cc\,\skew\left(\grad\u-P\right),\skew\left(\grad\u-P\right)\right\rangle _{\R^{3\times3}}}_{\textrm{invariant local anisotropic rotational elastic coupling}}\nonumber \\
 & \qquad+\underbrace{\frac{\mu\,L_{c}^{2}}{2}\left(\left\langle \mathbb{L}\,\sym\,\curl\,P,\sym\,\curl\,P\right\rangle _{\R^{3\times3}}+\left\langle \mathbb{L}_{c}\,\skew\,\curl\,P,\skew\,\curl\,P\right\rangle _{\R^{3\times3}}\right)}_{\textrm{ curvature}}\nonumber 
\end{align}
with
\[
\begin{cases}
\rho:\Omega\fr\R^{+} & \textrm{macro-inertia mass density},\\
\mathbb{J}_{\textrm{micro}}:\Sym\fr\Sym & \textrm{classical \ensuremath{4^{th}}order free micro-inertia density tensor},\\
\mathbb{T}:\Sym\fr\Sym & \textrm{classical \ensuremath{4^{th}}order gradient micro-inertia density tensor},\\
\mathbb{J}_{c},\mathbb{T}_{c}:\so\fr\so & \textrm{ \ensuremath{4^{th}}order coupling tensors with 6 independent components},\\
\\
\C_{e},\mathbb{C}_{\textrm{micro}},\mathbb{L}:\Sym\fr\Sym & \textrm{classical \ensuremath{4^{th}}order elasticity tensors with 21 independent components, }\\
\cc,\mathbb{L}_{c}:\so\fr\so & \textrm{ \ensuremath{4^{th}} order coupling tensors with 6 independent components},
\end{cases}
\]
where $L_{c}$ is the characteristic length of the relaxed micromorphic
model. We require that the bilinear forms induced by $\mathbb{J}_{\textrm{micro}},\mathbb{T},\C_{e},\mathbb{C}_{\textrm{micro}},\mathbb{L}$
are positive definite, while the bilinear forms induced by $\mathbb{J}_{c},\mathbb{T}_{c},\cc,\mathbb{L}_{c}$
are only required to be positive semi-definite. Expressions of the kinetic energy density involving the time derivative of $\nabla u$ and $P$ have been already proposed in literature as for example in \cite{hlavacek1975continuum}.
The equations of motion follow from the standard procedure and
are  given as a set of 3 coupled PDE-systems for $u$, $\sym\,P$
and $\skew\,P$:

\begin{figure}[H]
\centering{}%
\noindent\fbox{\begin{minipage}[t]{1\columnwidth - 2\fboxsep - 2\fboxrule}%
\vspace{-0.5cm}

\begin{align}
\rho\,\u_{,tt}-\textrm{Div}\left(\mathbb{T}\,\sym\,\nabla u_{,tt}\right)\nonumber \\
-\textrm{Div}\left(\mathbb{T}_{c}\,\skew\,\nabla u_{,tt}\right) & =\textrm{Div}\left(\ce\,\sym\left(\nabla\u-\P\right)+\cc\,\skew\left(\nabla\u-\P\right)\right),\label{eq:PDE system}\\
\mathbb{J}_{\textrm{micro}}\,\sym\,\P_{,tt} & =\ce\,\sym\left(\nabla\u-\P\right)-\mathbb{C}_{\textrm{micro}}\,\sym\,\P-\mu\,L_{c}^{2}\,\sym\,\curl\left(\mathbb{L}\,\sym\,\curl\,P+\mathbb{L}_{c}\,\skew\,\curl\,P\right),\nonumber \\
\mathbb{J}_{c}\,\skew\,\P_{,tt} & =\cc\,\skew\left(\nabla\u-\P\right)-\mu\,L_{c}^{2}\,\skew\,\curl\left(\mathbb{L}\,\sym\,\curl\,P+\mathbb{L}_{c}\,\skew\,\curl\,P\right).\nonumber 
\end{align}
\end{minipage}}
\end{figure}
\noindent The well posedness of this dynamical problem has been proven in \cite{owczarek2018nonstandard}.

\section{Plane wave propagation in anisotropic relaxed micromorphic media}

As it is known in the context of dynamical analysis, a particular
class of solutions of the system of partial differential equations
(\ref{eq:PDE system}) can be found considering the monochromatic
plane wave form for the involved kinematic fields, i.e., 
\begin{align}
\u\left(x,t\right) & =\widehat{u}\,e^{\gi\,\left(\left\langle \boldsymbol{k},x\right\rangle -\,\omega t\right)},\qquad P=\widehat{P}\,e^{\gi\,\left(\left\langle \boldsymbol{k},x\right\rangle -\,\omega t\right)},\qquad\boldsymbol{k}=k\,\widehat{\boldsymbol{k}},\label{eq:wave strucutre-1}
\end{align}
where $\widehat{u}=\left(\widehat{u}_{1},\widehat{u}_{2},\widehat{u}_{3}\right)$
is the so-called polarization vector in $\C^{3}$ , $\widehat{\boldsymbol{k}}=\left(k_{1},k_{2},k_{3}\right)\in\R^{3},\:\bigl\Vert\widehat{\boldsymbol{k}}\bigr\Vert=1$
is the direction of wave propagation and $\widehat{P}\in\C^{3\times3}$.
Under this hypothesis, when substituting (\ref{eq:wave strucutre-1})
into (\ref{eq:PDE system}), the search of solutions to (\ref{eq:PDE system})
turns into an algebraic problem. Indeed, the system of partial differential
equations (\ref{eq:PDE system}) turns into 
\begin{equation}
D\left(k,\omega\right)\cdot v=0,\label{eq:algebraic problem}
\end{equation}
where  $D\left(k,\omega\right)$ is a $12\times12$
matrix with complex-valued entries, whose components are functions of
the constitutive tensors\footnote{This means that the components $D_{ij}$ of the matrix $D$ are functions
$D_{ij}\left(k,\omega,\mathbb{T},\mathbb{T}_{c},\mathbb{J}_{\textrm{micro}},\mathbb{J}_{c},\ce,\cc,\cm,\mathbb{L},\mathbb{L}_{c},L_{c}\right)$.
In the following, we will explicitly state only the dependence on
$\left(k,\omega\right)$ if not differently specified. } and of $\left(k,\omega\right)$. Moreover, we set 
\begin{equation}
v=\left(\widehat{u}_{1},\widehat{P}^{D},\widehat{P}^{S},\widehat{u}_{2},\widehat{P}_{\left(12\right)},\widehat{P}_{\left[12\right]},\widehat{u}_{3},\widehat{P}_{\left(13\right)},\widehat{P}_{\left[13\right]},\widehat{P}_{\left(23\right)},\widehat{P}_{\left[23\right]},\widehat{P}^{V}\right)\in\mathbb{R}^{12},\label{vettore ampiezze}
\end{equation}
in which $\widehat{P}^{D},\widehat{P}^{S},\widehat{P}^{V},\widehat{P}_{\left(12\right)},\widehat{P}_{\left(13\right)},\widehat{P}_{\left(23\right)},\widehat{P}_{\left[12\right]},\widehat{P}_{\left[13\right]},\widehat{P}_{\left[23\right]},$
are defined, following \cite{madeo2015wave,dagostino2016panorama},
as
\begin{flalign*}
\widehat{P}^{D} & =\frac{2}{3}\widehat{P}_{11}-\frac{1}{3}\left(\widehat{P}_{22}+\widehat{P}_{33}\right), & \widehat{P}_{\left(rl\right)} & =\widehat{P}_{\left(lr\right)}=\frac{1}{2}\left(\widehat{P}_{rl}+\widehat{P}_{lr}\right)\;\textrm{if}\;r\neq l, & \widehat{P}_{\left[rl\right]} & =\frac{1}{2}\left(\widehat{P}_{rl}-\widehat{P}_{lr}\right)=-\widehat{P}_{\left[lr\right]},\\
\widehat{P}^{S} & =\frac{1}{3}\left(\widehat{P}_{11}+\widehat{P}_{22}+\widehat{P}_{33}\right), & \widehat{P}^{V} & =\widehat{P}_{22}-\widehat{P}_{33} & 2\,\widehat{\P}^{S}-\widehat{P}^{D} & =\widehat{\P}_{22}+\widehat{P}_{33}.
\end{flalign*}
Clearly, the algebraic problem (\ref{eq:algebraic problem}) admits
non-trivial solutions if and only if the determinant of the matrix
$D$ is zero. The equation $\det D=0$ allows to calculate the eigenvalues
$\omega=\omega\left(k\right)$. The curves $\omega=\omega\left(k\right)$
plotted in the $\left(\omega,k\right)$ plane are called \textbf{dispersion
curves}. Due to the complicated form of the components of the matrix $D\left(k,\omega\right)$
as a function of the constitutive parameters in the fully anisotropic
case, we will not explicitly write them here but, we can say that,
in general, the matrix $D\left(k,\omega\right)$ has the structure
$D\left(k,\omega\right)=C\,k^{2}\omega^{2}+A_{2}\,k^{2}+B_{2}\,\omega^{2}+A_{1}k+C_{0},$
where $C,A_{2},B_{2},A_{1},C_{0}$ are matrices in $\C^{12\times12}$
depending on the material parameters.

\subsection{Reduction to the 2D plain strain case}

The anisotropic relaxed micromorphic model (more
precisely, the model with tetragonal symmetry), supports purely in-plane
wave solutions in plain strain, provided we use the simplification
that $\mathbb{L}=\id$, $\mathbb{L}_{c}=\id$, since $\mathbb{L}\,\sym\curl\,P+\mathbb{L}_{c}\,\skew\curl\,P=\sym\,\curl\,P+\skew\,\curl\,P=\curl\,P$
in (\ref{eq:PDE system}) and then the operation $\curl\,\curl$ respects
the plain strain format, i.e., 
\begin{equation}
\curl\,\curl\begin{pmatrix}\star & \star & 0\\
\star & \star & 0\\
0 & 0 & 0
\end{pmatrix}=\begin{pmatrix}\star & \star & 0\\
\star & \star & 0\\
0 & 0 & 0
\end{pmatrix}.
\end{equation}
We take advantage of this simplification and we will be interested
(for simplicity of the computational task) only in the study of the
wave propagation with $\widehat{\boldsymbol{k}}$ in the plane $\left(x_{1},x_{2},0\right)$.
In this way, setting the amplitudes out of the plane, $\widehat{u}_{3},\widehat{P}_{13},\widehat{P}_{23},\widehat{P}_{33},\widehat{P}_{31},\widehat{P}_{32}$,
equal to zero, the vector of the unknown amplitudes $v$ given in
(\ref{vettore ampiezze}) reduces to\footnote{Note that once $\widehat{P}^{D}$ and $\widehat{P}^{S}$ are known
then $\widehat{P}^{V}$ is automatically determined and, in general,
not vanishing.} 
\[
\widetilde{v}=\left(\widehat{u}_{1},\widehat{P}^{D},\widehat{P}^{S},\widehat{u}_{2},\widehat{P}_{\left(12\right)},\widehat{P}_{\left[12\right]}\right).
\]
The system of algebraic equations (\ref{eq:algebraic problem}) simplifies
to $\widetilde{D}\left(k,\omega\right)\cdot\widetilde{v}=0,\label{reduced algebraic system}$. The dispersion curves
$\omega=\omega\left(k\right)$ are the solutions of the algebraic equation $\det\widetilde{D}\left(k,\omega\right)=0.$

\section{Bloch-Floquet analysis}

The Bloch-Floquet theorem is routinely used for computing the dispersion properties of liner elastic periodic structures, as well as for computing the wave modes and group velocities. 

The framework allows reducing computational costs  by modelling of a representative cell, at the same time providing a rigorous and well-posed spectral problem representing dispersion of waves in unbounded media.

The Bloch-Floquet theorem is tailored for the representation of solution of systems of PDEs with periodic coefficients. Considering an infinite periodic elastodynamic problem, the time harmonic dynamical equilibrium of the system is governed by
\begin{equation}\label{Bloch1}
\rho(x)\,\omega^2\,u(x)+\textrm{Div}\left[ \C(x)\,\sym\nabla u(x)\right] =0,\qquad\forall x\in\R^3
\end{equation}
where $u:\R^3\fr\C^3$ is the displacement field and the mass density $\rho(x)$ and the elasticity tensor $\C(x)$ are periodic functions. We assume also that $\rho(x)$ is bounded from below by a positive constant and $\C(x)$ is uniformly elliptic. Introducing the differential operator
$$
\mathcal{A}:\mathcal{D}\left(\mathcal{A}\right)\subseteq L^{2}\left(\mathbb{R}^{3},\mathbb{C}^{3}\right)\longrightarrow L^{2}\left(\mathbb{R}^{3},\mathbb{C}^{3}\right),\qquad u\mapsto\mathcal{A}\left[u\right]\defi-\frac{1}{\rho\left(x\right)}\,\textrm{Div}_{x}\left[\C\left(x\right)\cdot\textrm{sym}\,\nabla u\left(x,t\right)\right],
$$
and setting $\lambda=\omega^2$, the differential problem \eqref{Bloch1} can be reformulated as an eigenvalues problem for the operator $\mathcal{A}$. The Bloch-Floquet analysis allows us to decompose this spectral problem in a family of easier spectral problems parametrized by the elements $k$ of the Brilloin zone $\Lambda$. More precisely, let us introduce the shifting operators
$$
u_{k}\mapsto\mathcal{A}_k\left[u_{k}\right]\defi-\frac{1}{\rho(x)}\,\textrm{Div}_{x}\left[\C\left(x\right)\cdot\textrm{sym}\left(\nabla u_{k}+\gi\,k\otimes u_{k}\right)\right]-\frac{1}{\rho(x)}\left[\C\left(x\right)\cdot\textrm{sym}\left(\nabla u_{k}+\gi\,k\otimes u_{k}\right)\right]\cdot \gi\,k,
$$
whose domains are vector subspaces of the space $H^1_{\textrm{per}}(V,\C^3)$ of the periodic Sobolev functions defined on the unit cell $V$. It can be shown that the spectrum of the shifting operators is discrete and it accumulates at infinity, i.e., given a $k\in\Lambda$ we have that
$$
\sigma\left( \mathcal{A}_k\right)=\left\lbrace \lambda_n(k) \right\rbrace_{n=1}^\infty\subseteq\R\;:\;\lim_{n\fr\infty}\lambda_n(k)=+\infty.  
$$
In this way, Bloch-Floquet theory allows to reduce the wave propagation problem in a periodic medium to a unit-cell with appropriate boundary conditions (similar to the determination of the effective stiffness tensor on the unit-cell using periodic boundary conditions).
In order to obtain numerically the spectra $\sigma\left( \mathcal{A}_k\right)$, we rewrite the problem in the weak form introducing the bilinear forms
\begin{align*}
\mathscr{B}_k:H^1_{\textrm{per}}(V,\C^3)\times H^1_{\textrm{per}}(V,\C^3)&\fr\C,\\\mathscr{B}_k\left[u_{k},v_{k}\right]&\defi\int_{V}\left\langle \C\left(x\right)\cdot\textrm{sym}\left(\nabla u_{k}+\gi\,k\otimes u_{k}\right),\overline{\textrm{sym}\left(\nabla v_{k}+\gi\,k\otimes v_{k}\right)}\right\rangle dm,
\end{align*}
Using standard FEM computation, the problem is projected to suitable finite-dimensional vector subspaces allowing us to find the first eigenvalues of the accounted differential operators. The main statement of the Bloch-Floquet theorem consists in showing that the full spectrum of $\mathcal{A}$ is the closure of the union of all the spectra of the operators $\mathcal{A}_k$.
\begin{figure}[H]
	\begin{centering}
		\includegraphics[scale=0.40]{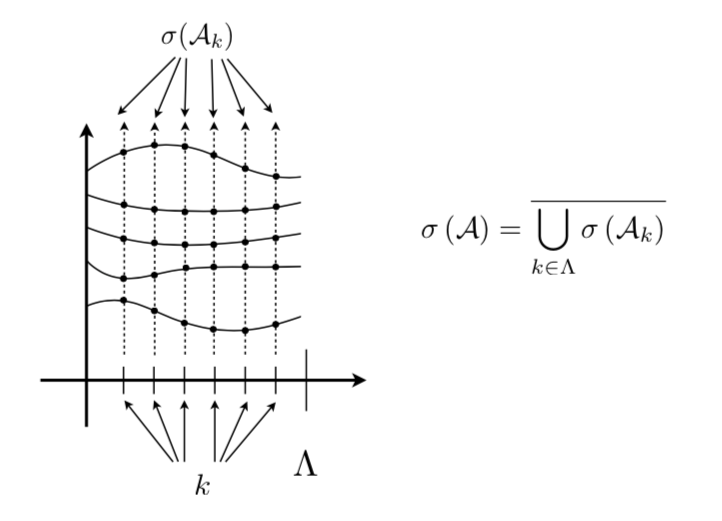}\caption{Spectral decomposition for the eigenvalues problem for periodic coefficients differential operators (Bloch-Floquet theorem).}
		\par\end{centering}
\end{figure}

\section{\label{sec:metamaterial}Unit-cell and numerical simulations via Bloch-Floquet analysis}

\begin{figure}[h]
\begin{centering}
\begin{tabular}[t]{ccc}
\includegraphics[scale=0.35]{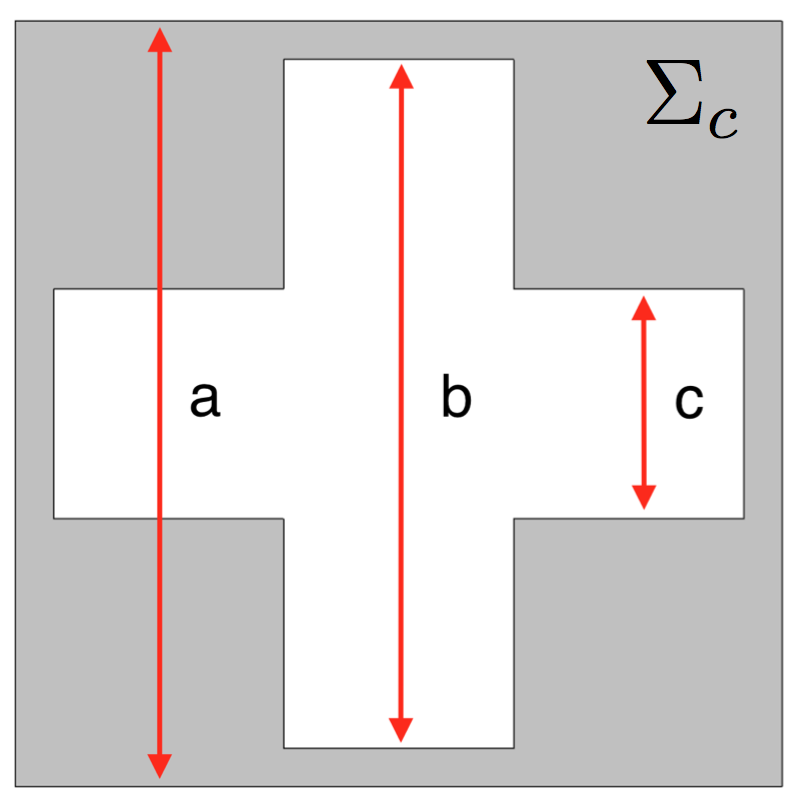} & \includegraphics[scale=0.4]{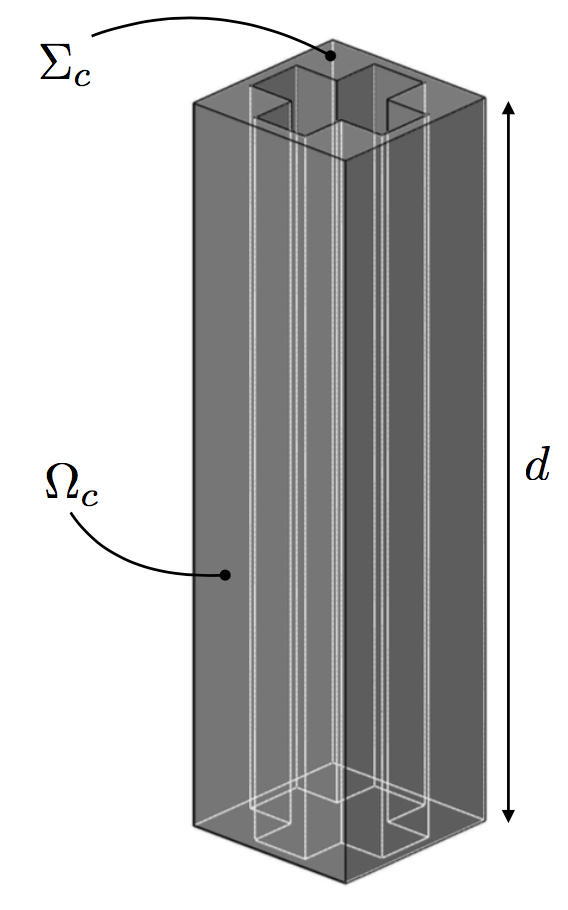} & \includegraphics[scale=0.35]{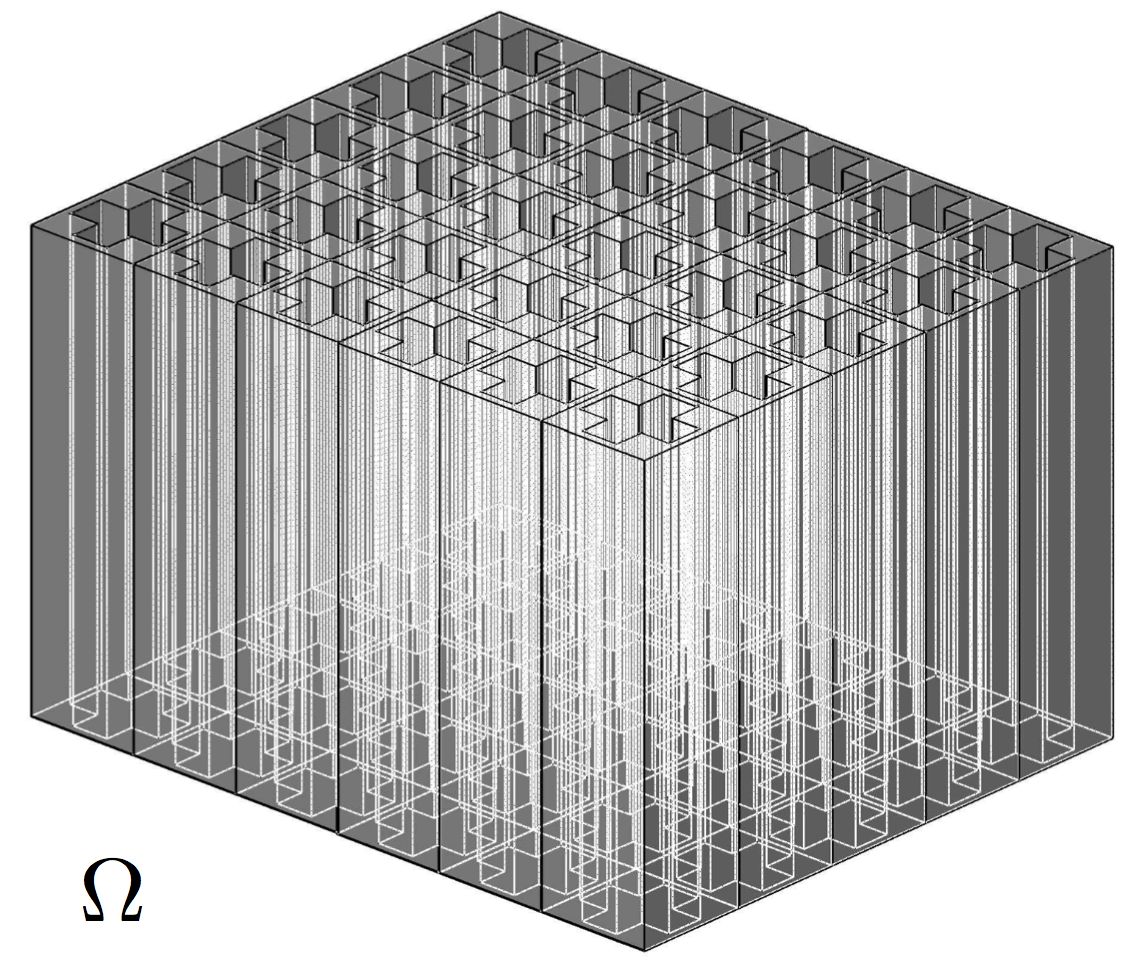}\tabularnewline
(a) & (b) & (c)\tabularnewline
\end{tabular}
\par\end{centering}
\caption{\label{fig:Microstructure-implemented-in}Microstructure implemented
in COMSOL$^{\circledR}$: (a) Plane cell $\Sigma_{c}$, (b) 3D tetragonal
cell $\Omega_{c}$, (c) infinite periodic medium $\Omega$.}
\end{figure}

In this Section, we perform some discrete numerical simulations of
wave dispersion in a precise metamaterial which will be further used
to suitably show how the proposed relaxed micromorphic model can describe
its (effective) homogenized behavior. Chosen the microstructure, we
perform a standard Bloch-Floquet analysis of the wave propagation in the generated
periodic infinite medium thanks to the FEM code COMSOL$^{\circledR}$.
This kind of analysis can be easily implemented using the Bloch-Floquet
boundary conditions which are built in the code. The microstructure
$\Omega_{c}$ (see Fig.\ref{fig:Microstructure-implemented-in}(b))
we account for is realized as follows: given the plane structure $\Sigma_{c}$
shown in Fig.\ref{fig:Microstructure-implemented-in}(a), with dimensions
specified in Table \ref{parametri microstruttura}, we define $\Omega_{c}=\Sigma_{c}\times\left[0,1\right]$
in which the unit is in meters. The grey region of $\Sigma_{c}$ is
filled by aluminum while the white one is empty. The group symmetry
of the introduced microstructure is the tetragonal one (the generated solid is invariant under the
action of the discrete subgroup\footnote{$\mathbb{D}_{4}$ is the dihedral group of order 4. It counts 8 elements.}
$\mathbb{D}_{4}$ of $\textrm{SO}\!\left(3\right)$). 

The geometric dimensions and the mechanical parameters (Young's modulus
and Poisson's ratio) of the presented microstructure are given in
Table \ref{parametri microstruttura}.

\begin{figure}[H]
\begin{centering}
\begin{tabular}{cc}
\includegraphics[scale=0.6]{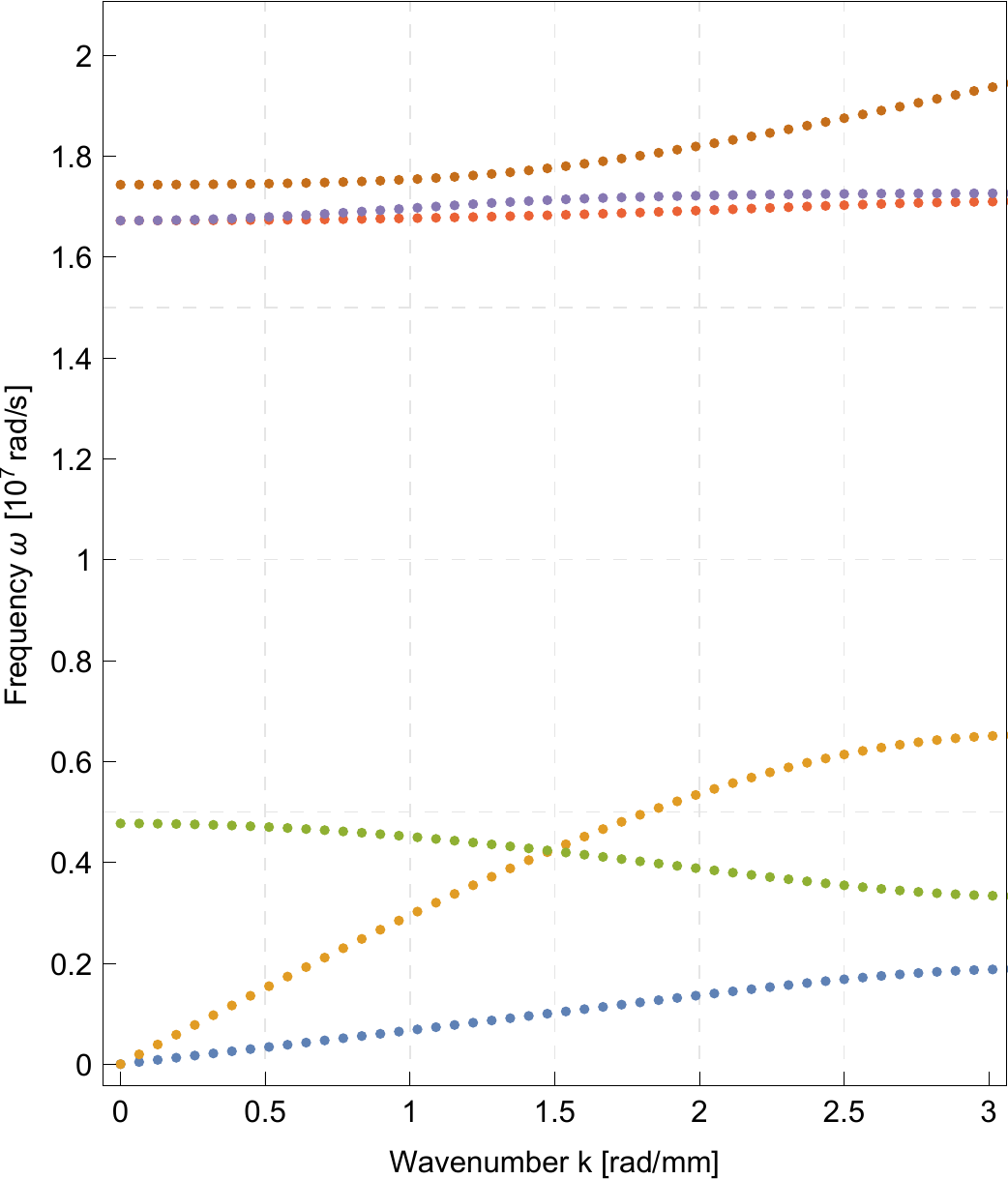} & \includegraphics[scale=0.6]{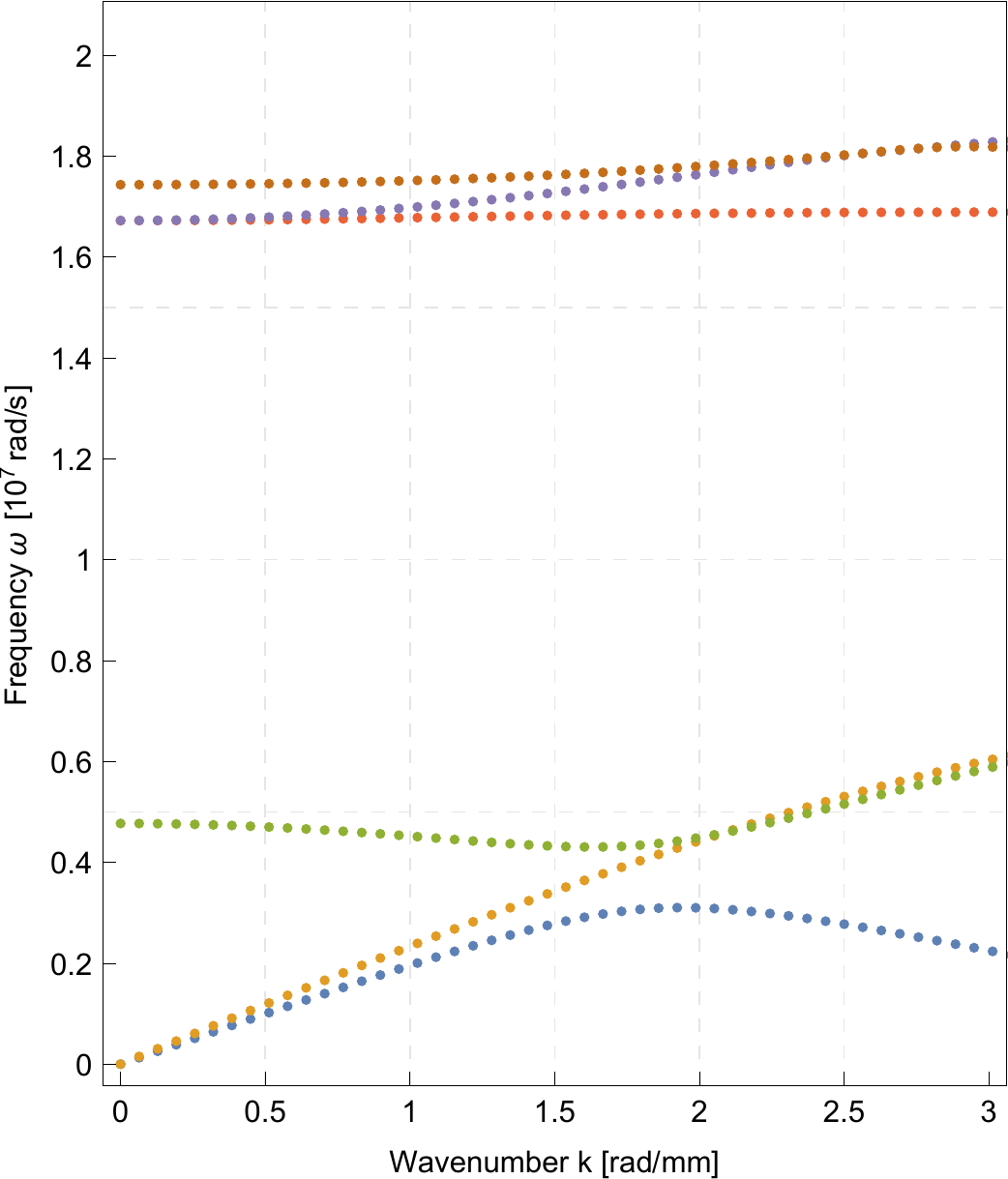}\tabularnewline
(a) & (b)\tabularnewline
\end{tabular}
\par\end{centering}
\caption{\label{fig:Comsol model} COMSOL$^{\circledR}$ model (plane strain
Bloch-Floquet analysis). In (a) we plot the dispersion branches for
$\widehat{\boldsymbol{k}}=\left(1,0,0\right)$ (horizontal wave propagation)
and in (b) for $\widehat{\boldsymbol{k}}=\left(\sqrt{2}/2,\sqrt{2}/2,0\right)$
(wave propagation at $45\text{\textdegree}$). We observe a band-gap.}

\end{figure}

\begin{table}[H]
\begin{centering}
\begin{tabular}{cccc}
$a$ & $b$ & $c$ & $d$\tabularnewline[1mm]
\hline 
\noalign{\vskip1mm}
$\left[\mathrm{mm}\right]$ & $\left[\mathrm{mm}\right]$ & $\left[\mathrm{mm}\right]$ & $\left[\mathrm{m}\right]$\tabularnewline[1mm]
\hline 
\hline 
\noalign{\vskip1mm}
$1$ & $0.9$ & $0.3$ & $1$\tabularnewline[1mm]
\end{tabular}$\qquad\qquad$%
\begin{tabular}{cccc}
$E$ & $\nu$ & $\mu$ & $\lambda$\tabularnewline[1mm]
\hline 
\noalign{\vskip1mm}
$\left[\mathrm{GPa}\right]$ & $-$ & $\left[\mathrm{GPa}\right]$ & $\left[\mathrm{GPa}\right]$\tabularnewline[1mm]
\hline 
\hline 
\noalign{\vskip1mm}
$70$ & $0.33$ & $26.32$ & $51.08$\tabularnewline[1mm]
\end{tabular}
\par\end{centering}
\caption{\label{parametri microstruttura}Geometry of the unit-cell (Fig.\,\ref{fig:Microstructure-implemented-in})
and elastic parameters of Aluminum.}
\end{table}
We can now determine the apparent density $\rho$ of the unit-cell
$\Omega_{c}$. In order to do this, denoting with $V_{\textrm{al}}$
the volume occupied by the aluminum in the unit-cell ($55$\%), we
find that 
\[
M_{\textrm{al}}=\rho_{\textrm{al}}V_{\textrm{al}}=2.7\times10^{3}\times5.5\times10^{-7}\left[\frac{\mathrm{kg}}{\mathrm{m}^{3}}\right]\left[\mathrm{m}^{3}\right]=14.85\times10^{-4}\left[\mathrm{kg}\right],
\]
where $M_{\textrm{al}}$ is the mass of the volume occupied by the
aluminum and $\rho_{\textrm{al}}=2.7\times10^{3}\left[\frac{\mathrm{kg}}{\mathrm{m}^{3}}\right]$
is the aluminum mass density. Since the volume of the unit-cell
is $\mathit{Vol}\left(\Omega_{c}\right)=10^{-6}\left[\mathrm{m}^{3}\right]$
we find that
\begin{equation}
\rho=\frac{M_{\textrm{al}}}{\mathit{Vol}\left(\Omega_{c}\right)}\left[\frac{\mathrm{kg}}{\mathrm{m}^{3}}\right]=1485\left[\frac{\mathrm{kg}}{\mathrm{m}^{3}}\right].
\end{equation}
When fixing the parameters of the unit-cell as in Table \ref{parametri microstruttura}
and when performing a Bloch-Floquet analysis on the considered periodic
structure, the dispersion curves shown in Fig.\ref{fig:Comsol model}
are obtained. 
These curves for the two directions will be used in Section \ref{sec:Fitting}
to fit the dynamical parameters of the relaxed micromorphic model. 

\section{The tetragonal case in the relaxed micromorphic model}

In this Section we are able to illustrate one of the advantages of using
the proposed relaxed micromorphic model for the description of the
homogenized mechanical behavior of anisotropic metamaterials. Indeed,
classical elastic tensors of linear elasticity can be used once the
symmetry class of the material is identified. This avoids unnecessary
complexifications related to the study of the symmetry classes of
higher order tensors as happens, e.g., in gradient elasticity (see,
e.g., \cite{auffray2013algebraic,Rosi2017validity}).

Since the crystallographic symmetry group of $\Omega_{c}$
is $\mathbb{D}_{4}$, we specify the general anisotropic model of
a continuum which has the tetragonal symmetry property. This means
that all involved structural tensors $\mathbb{C}$ have to respect
the invariance condition
\begin{equation}
Q_{ai}Q_{bj}Q_{ch}Q_{dk}\mathbb{C}{}_{abcd}=\mathbb{C}{}_{ijhk},\qquad\forall Q\in\mathbb{D}_{4}.
\end{equation}
We express the constitutive tensors as $6\times6$ matrices with tilde in Voigt notation.
In the considered tetragonal case, the matrices corresponding to the
considered tensors  have the structure (see, e.g., \cite{barbagallo2016transparent}):
\begingroup
\allowdisplaybreaks
\begin{align*}
\widetilde{\mathbb{C}}_{e} & =\begin{pmatrix}2\me+\le & \le & \le^{*} & 0 & 0 & 0\\
\le & 2\me+\le & \le^{*} & 0 & 0 & 0\\
\le^{*} & \le^{*} & (\widetilde{\mathbb{C}}_{e})_{33} & 0 & 0 & 0\\
0 & 0 & 0 & (\widetilde{\mathbb{C}}_{e})_{44} & 0 & 0\\
0 & 0 & 0 & 0 & (\widetilde{\mathbb{C}}_{e})_{44} & 0\\
0 & 0 & 0 & 0 & 0 & \me^{*}
\end{pmatrix}, & \widetilde{\mathbb{C}}_{c} & =\begin{pmatrix}4\mc^{*} & 0 & 0\\
0 & 4\mc^{*} & 0\\
0 & 0 & 4\mc
\end{pmatrix},\\
\\
\widetilde{\mathbb{L}} & =\begin{pmatrix}2\a_{1}+\a_{3} & \a_{3} & \a_{3}^{*} & 0 & 0 & 0\\
\a_{3} & 2\a_{1}+\a_{3} & \a_{3}^{*} & 0 & 0 & 0\\
\a_{3}^{*} & \a_{3}^{*} & \widetilde{\mathbb{L}}{}_{33} & 0 & 0 & 0\\
0 & 0 & 0 & \widetilde{\mathbb{L}}{}_{44} & 0 & 0\\
0 & 0 & 0 & 0 & \widetilde{\mathbb{L}}{}_{44} & 0\\
0 & 0 & 0 & 0 & 0 & \alpha_{1}^{*}
\end{pmatrix}, & \widetilde{\mathbb{L}}_{c} & =\begin{pmatrix}4\alpha_{2}^{*} & 0 & 0\\
0 & 4\alpha_{2}^{*} & 0\\
0 & 0 & 4\alpha_{2}
\end{pmatrix},\\
\\
\widetilde{\mathbb{T}} & =\begin{pmatrix}2\overline{\eta}_{1}+\overline{\eta}_{3} & \overline{\eta}_{3} & \overline{\eta}_{3}^{*} & 0 & 0 & 0\\
\overline{\eta}_{3} & 2\overline{\eta}_{1}+\overline{\eta}_{3} & \overline{\eta}_{3}^{*} & 0 & 0 & 0\\
\overline{\eta}_{3}^{*} & \overline{\eta}_{3}^{*} & \widetilde{\mathbb{T}}{}_{33} & 0 & 0 & 0\\
0 & 0 & 0 & \widetilde{\mathbb{T}}{}_{44} & 0 & 0\\
0 & 0 & 0 & 0 & \widetilde{\mathbb{T}}{}_{44} & 0\\
0 & 0 & 0 & 0 & 0 & \overline{\eta}_{1}^{*}
\end{pmatrix}, & \widetilde{\mathbb{T}}_{c} & =\begin{pmatrix}4\overline{\eta}_{2}^{*} & 0 & 0\\
0 & 4\overline{\eta}_{2}^{*} & 0\\
0 & 0 & 4\overline{\eta}_{2}
\end{pmatrix},
\end{align*}
\endgroup

\begin{align}
\widetilde{\mathbb{J}}_{\textrm{micro}} & =\begin{pmatrix}2\eta_{1}+\eta_{3} & \eta_{3} & \eta_{3}^{*} & 0 & 0 & 0\\
\eta_{3} & 2\eta_{1}+\eta_{3} & \eta_{3}^{*} & 0 & 0 & 0\\
\eta_{3}^{*} & \eta_{3}^{*} & (\widetilde{\mathbb{J}}_{\textrm{micro}})_{33} & 0 & 0 & 0\\
0 & 0 & 0 & (\widetilde{\mathbb{J}}_{\textrm{micro}})_{44} & 0 & 0\\
0 & 0 & 0 & 0 & (\widetilde{\mathbb{J}}_{\textrm{micro}})_{44} & 0\\
0 & 0 & 0 & 0 & 0 & \eta_{1}^{*}
\end{pmatrix}, & \widetilde{\mathbb{J}}_{c} & =\begin{pmatrix}4\eta_{2}^{*} & 0 & 0\\
0 & 4\eta_{2}^{*} & 0\\
0 & 0 & 4\eta_{2}
\end{pmatrix},
\end{align}
\begin{align*}
\widetilde{\mathbb{C}}_{\textrm{micro}} & =\begin{pmatrix}2\,\mu_{\textrm{micro}}+\lambda_{\textrm{micro}} & \lambda_{\textrm{micro}} & \lambda_{\textrm{micro}}^{*} & 0 & 0 & 0\\
\lambda_{\textrm{micro}} & 2\,\mu_{\textrm{micro}}+\lambda_{\textrm{micro}} & \lambda_{\textrm{micro}}^{*} & 0 & 0 & 0\\
\lambda_{\textrm{micro}}^{*} & \lambda_{\textrm{micro}}^{*} & (\widetilde{\mathbb{C}}_{\textrm{micro}})_{33} & 0 & 0 & 0\\
0 & 0 & 0 & (\widetilde{\mathbb{C}}_{\textrm{micro}})_{44} & 0 & 0\\
0 & 0 & 0 & 0 & (\widetilde{\mathbb{C}}_{\textrm{micro}})_{44} & 0\\
0 & 0 & 0 & 0 & 0 & \mu_{\textrm{micro}}^{*}
\end{pmatrix}.
\end{align*}

\begin{rem}
In the considered plane strain 2D case (no
micro and macro motion in the 3-direction) some of the components
of the elastic tensors do explicitly appear neither in the PDEs
(\ref{eq:PDE system}) nor in the algebraic system (\ref{reduced algebraic system}).
This is equivalent to say that, in the considered 2D tetragonal case,
the only active components of the involved elastic tensors can be
identified as follows:

\begin{align}
 & \widetilde{\mathbb{C}}_{e}=\begin{pmatrix}2\me+\le & \le & \star & 0 & 0 & 0\\
\le & 2\me+\le & \star & 0 & 0 & 0\\
\star & \star & \star & 0 & 0 & 0\\
0 & 0 & 0 & \star & 0 & 0\\
0 & 0 & 0 & 0 & \star & 0\\
0 & 0 & 0 & 0 & 0 & \me^{*}
\end{pmatrix}, &  & \widetilde{\mathbb{C}}_{c}=\begin{pmatrix}\star & 0 & 0\\
0 & \star & 0\\
0 & 0 & 4\mc
\end{pmatrix}.
\end{align}
Thus, we may arrange the elasticity tensor for our purpose as 
\begin{equation}
\begin{pmatrix}2\mu_{e}+\lambda_{e} & \lambda_{e} & 0\\
\lambda_{e} & 2\mu_{e}+\lambda_{e} & 0\\
0 & 0 & \mu_{e}^{*}
\end{pmatrix}\label{eq:reduction}
\end{equation}
acting only on $\left(\varepsilon_{11},\varepsilon_{22},\varepsilon_{12}\right)$.
A similar reducibility holds for the other involved tensors. 
\end{rem}

\section{\label{sec:Fitting}Fitting material parameters and analysis of dispersion
curves}

In this Section we will show how:
\begin{itemize}
\item thanks to the general anisotropic micro-macro homogenization formula
developed in \cite{barbagallo2016transparent} it is possible to establish
a functional dependence between the components of the tensors $\mathbb{C}_{e}$
and $\mathbb{C}_{\textrm{micro}}$ appearing in the relaxed micromorphic
model and those of the elastic macroscopic elasticity tensor $\mathbb{C}_{\textrm{macro}}$
of the effective (Cauchy) relaxed micromorphic limit model when considering
the tetragonal case. Since $\cm$ and $\cM$ are known from static
arguments, as shown in \cite{d2019Identification}, $\ce$ can be readily computed,
\item the calculation of the cut-off frequencies provides the possibility
to obtain four extra relations between the micro elastic parameters
and the micro inertia terms, thus finally allowing the computation
of the micro inertiae and of the Cosserat couple modulus $\mc$.
\end{itemize}

\subsection{Micro-macro homogenization formula}

We now specialize the tensorial micro-macro homogenization formulas
obtained in \cite{barbagallo2016transparent,neff2007geometrically,neff2004material}
for the general anisotropic framework to the tetragonal case. To this
aim, we start counting the parameters of the relaxed micromorphic
model for the tetragonal case, which are
$
\left(\me,\le,\me^{*},\mc,\mu_{\textrm{micro}},\lambda_{\textrm{micro}},\mu_{\textrm{micro}}^{*},\mu_{\textrm{macro}},\lambda_{\textrm{macro}},\mu_{\textrm{macro}}^{*},L_{c},\a_{1},\a_{2},\a_{3},\a_{1}^{*}\right)
$
for the potential part of the energy and
$
\left(\rho,\eta_{1},\eta_{2},\eta_{3},\eta_{1}^{*},\overline{\eta}_{1},\overline{\eta}_{2},\overline{\eta}_{3},\overline{\eta}_{1}^{*}\right)
$
for the kinetic one, then we use the fundamental homogenization formula
found in \cite{barbagallo2016transparent}. In \cite{barbagallo2016transparent}
it is shown that in the limit $L_{c}\fr0$ it is possible to homogenize
the relaxed micromorphic model to a Cauchy one whose elastic (macroscopic)
stiffness $\mathbb{C}_{\textrm{macro}}$ is linked to the relaxed
micromorphic material parameters, $\mathbb{C}_{\textrm{micro}},\ce$
by the relation
\begin{align}\label{eq:homog formula}
\mathbb{C}_{\textrm{macro}} & =\mathbb{C}_{\textrm{micro}}\left(\mathbb{C}_{\textrm{micro}}+\mathbb{C}_{e}\right)^{-1}\mathbb{C}_{e}
&\Longleftrightarrow& &\mathbb{C}_{e} & =\mathbb{C}_{\textrm{micro}}\left(\mathbb{C}_{\textrm{micro}}-\mathbb{C}_{\textrm{macro}}\right)^{-1}\mathbb{C}_{\textrm{macro}},
\end{align}
which, in our tetragonal case, leads to the identities
\begin{align}
\mu_{\textrm{macro}} & =\frac{\me\,\mu_{\textrm{micro}}}{\me+\mu_{\textrm{micro}}}, & \!\!\!\!\!2\mu_{\textrm{macro}}+3\lambda_{\textrm{macro}} & =\frac{\left(2\me+3\le\right)\left(2\mu_{\textrm{micro}}+3\lambda_{\textrm{micro}}\right)}{2\left(\me+\mu_{\textrm{micro}}\right)+3\left(\le+\lambda_{\textrm{micro}}\right)}, & \!\!\!\!\!\mu_{\textrm{macro}}^{*} & =\frac{\me^{*}\,\mu_{\textrm{micro}}^{*}}{\me^{*}+\mu_{\textrm{micro}}^{*}},\nonumber \\
 &  &  & \Longleftrightarrow\label{eq:mue-muh}\\
\me & =\frac{\mu_{\textrm{macro}}\,\mu_{\textrm{micro}}}{\mu_{\textrm{micro}}-\mu_{\textrm{macro}}}, & \!\!\!\!\!2\me+3\le & =\frac{\left(2\mu_{\textrm{macro}}+3\lambda_{\textrm{macro}}\right)\left(2\mu_{\textrm{micro}}+3\lambda_{\textrm{micro}}\right)}{\left(2\mu_{\textrm{micro}}+3\lambda_{\textrm{micro}}\right)-\left(2\mu_{\textrm{macro}}+3\lambda_{\textrm{macro}}\right)}, & \!\!\!\!\!\me^{*} & =\frac{\mu_{\textrm{macro}}^{*}\,\mu_{\textrm{micro}}^{*}}{\mu_{\textrm{micro}}^{*}-\mu_{\textrm{macro}}^{*}}.\nonumber 
\end{align}
Equations (\ref{eq:mue-muh}) give the explicit relations between
the parameters of the tetragonal relaxed micromorphic model and the
corresponding macroscopic parameters of the Cauchy model seen as a
limiting case of the relaxed micromorphic model when $L_{c}\fr0$, see \cite{d2019Identification} for further explanations.

The importance of this micro-macro homogenization formula can hardly
be overestimated. Indeed, it allows for calibrating the a priori unknown
material parameters of the linear relaxed micromorphic model $\ce$
against the in principle known and measurable macroscopic response
$\cM$ and microscopic response $\cm$. In our case, $\cM$ has been
obtained via numerical homogenization of periodic media (see \cite{d2019Identification}). On the other hand, $\cM$ could equivalently be 
determined from a comparison with the Bloch-Floquet analysis (see also \cite{zhikov2016operator} for a good review about homogenization techniques in this context). More precisely,
it is well known that the macroscopic parameters $\mu_{\textrm{macro}},\lambda_{\textrm{macro}}$
and $\mu_{\textrm{macro}}^{*}$ can be directly related to the slopes
of the acoustic branches of a Cauchy continuum with tetragonal symmetry,
and this will allow the determination of the macro parameters for
the given metamaterial. We have checked that the macro parameters obtained
with the two methods turn out to be the same. 

The clear physical interpretation of the micro stiffnesses $\mathbb{C}_{\textrm{micro}}$,
on the other hand, is more complicated. Indeed, we know that such
micro stiffnesses must be related to the mechanical properties of
the unit-cell and true experimental static tests should be run on
a specimen composed by a single unit-cell in order to obtain their
values. As we explained in \cite{d2019Identification}, different stiffnesses
of the unit-cell can be obtained when changing the representative
unit-cell for the same metamaterial. However, we established rigorously
that the elastic parameters of the stiffest response on the micro-scale
must be chosen as $\cm$.

Once the macro and micro parameters $\cM$ and $\cm$ have been determined,
the micro-macro homogenization formula (\ref{eq:homog formula}) allows
us to uniquely determine the stiffness $\ce$ which realizes the transition
between the macro and micro scale. In particular, for the considered
tetragonal metamaterial, and considering the values of $\cM$ and
$\cm$ given in \cite{d2019Identification}, the values of $\ce$ given in Table\,\ref{C_e}
can be easily computed through the homogenization formula (\ref{eq:mue-muh}).

\begin{table}[H]
\begin{centering}
\begin{tabular}{ccc}
$\le$ & $\mu_{e}$ & $\me^{*}$\tabularnewline[1mm]
\hline 
\noalign{\vskip1mm}
$\left[\textrm{GPa}\right]$ & $\left[\textrm{GPa}\right]$ & $\left[\textrm{GPa}\right]$\tabularnewline[1mm]
\hline 
\hline 
\noalign{\vskip1mm}
$-\,0.77$ & $17.34$ & $0.67$\tabularnewline[1mm]
\end{tabular}
\par\end{centering}
\caption{\label{C_e}Obtained numerical values for the parameters related to
the mesoscopic transition scale. The tensor $\protect\ce$ remains positive definite.}
\end{table}

\subsection{Cut offs for the optic branches}

Only the Cosserat couple modulus $\mc$ remains to be determined as far as
the purely elastic parameters are concerned. On the other hand, all
the micro inertiae $\eta_{1},\eta_{1}^{*},\eta_{2},\eta_{3},\overline{\eta}_{1},\overline{\eta}_{1}^{*},\overline{\eta}_{2},\overline{\eta}_{3}$
appearing in the kinetic energy (\ref{kin energy}) remain to be determined
as well. To this aim we consider the so-called cut-off frequencies.
Solving the equation $\det\widetilde{D}=0$ imposing that $k=0$,
we find the following characteristic frequencies:

\begin{align}
\omega_{r} & =\sqrt{\frac{\mc}{\eta_{2}}}, & \omega_{s} & =\sqrt{\frac{\me+\mh}{\eta_{1}}}, & \omega_{s}^{*} & =\sqrt{\frac{\me^{*}+\mh^{*}}{\eta_{1}^{*}}}, & \omega_{p} & =\sqrt{\frac{\me+\mh+\le+\lh}{\eta_{1}+\eta_{3}}}.\label{cut-offs}
\end{align}
Such characteristic frequencies correspond to the starting point $\left(k=0\right)$
of the dispersion curves and are known as cut-off frequencies. They
are independent of the wave direction.

The simple fact of imposing that such characteristic frequencies are
equal to the numerical values of the cut-offs calculated via the Bloch-Floquet
analysis (see Fig.(\ref{fig:Comsol model})) allows us to establish
specific relations for computing some of the parameters of the relaxed
micromorphic model which are still free. In particular, as we will
show in the next Section, the last three formulas of equation (\ref{cut-offs})
enable us to compute $\eta_{1},\eta_{1}^{*}$ and $\eta_{3}$. 

\subsection{\label{subsec:Fitting-and-dispersion}Fitting of the parameters on
the dispersion curves}

\begin{figure}[H]
\begin{centering}
\begin{tabular}{cc}
\includegraphics[scale=0.5]{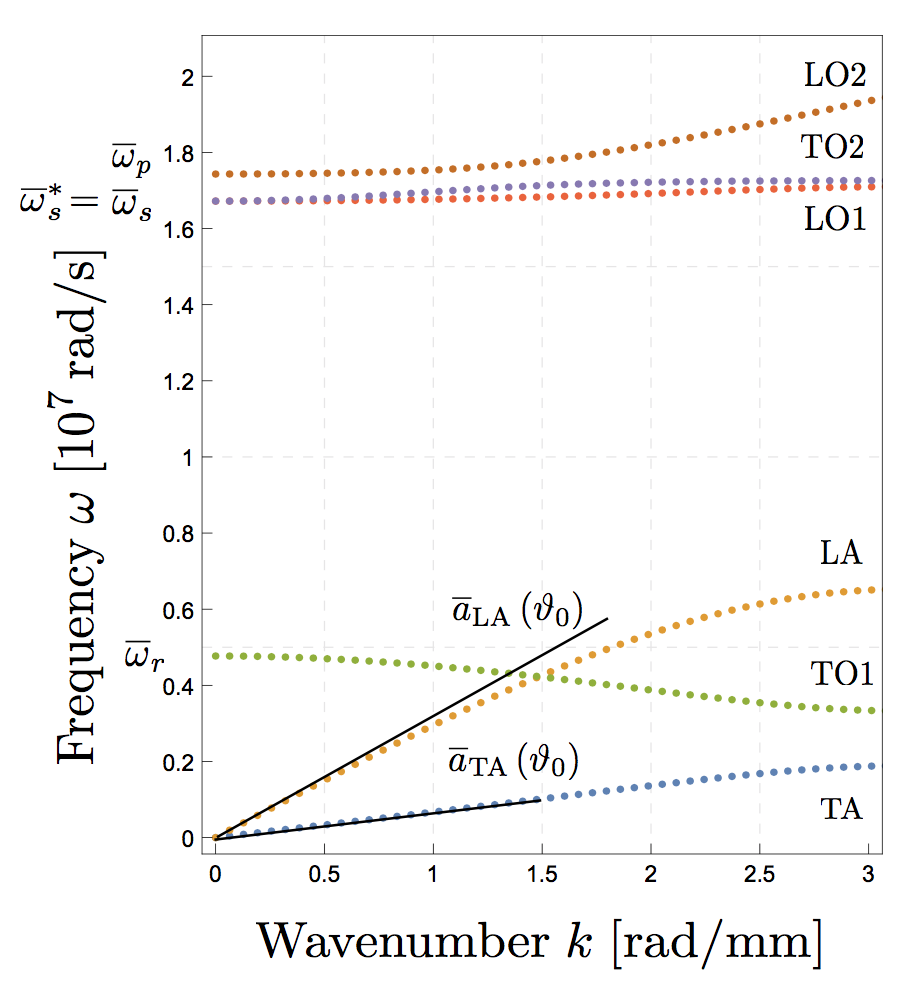} & \includegraphics[scale=0.5]{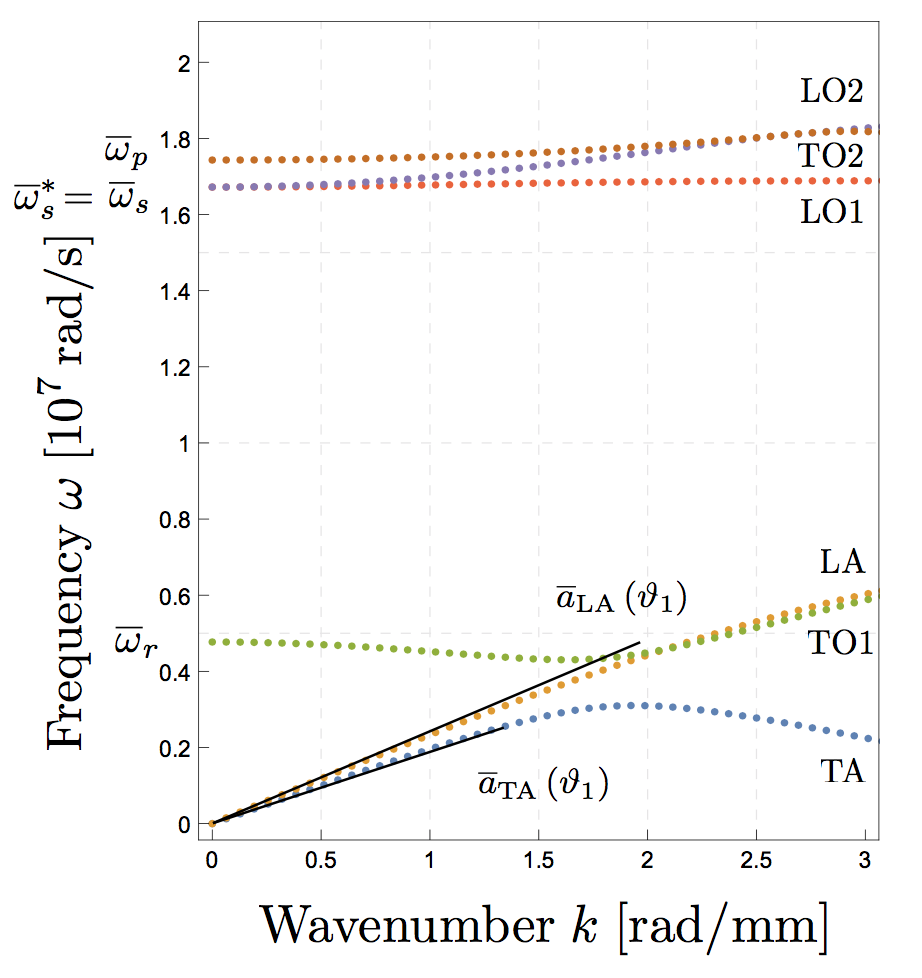}\tabularnewline
\end{tabular}
\par\end{centering}
\caption{\label{fig:Dispersion-curves-of}Dispersion curves of the selected
metamaterial (for the two directions $\vartheta=\vartheta_0=0$ and $\vartheta=\vartheta_1=\pi/4$) and identification of the key numerical quantities needed
for the fitting procedure. The cut off frequency $\omega_{r},\omega_{s},\omega_{s}^{*}$
and $\omega_{p}$ are direction independent, while the tangents to
the acoustic curves $\overline{a}_{\textrm{TA}}$ and $\overline{a}_{\textrm{LA}}$
depend on the direction of wave propagation.}

\end{figure}

In this Section we show the fitting procedure that we used to calibrate
the remaining free parameters of our relaxed micromorphic model on
the metamaterial introduced in Section \ref{sec:metamaterial}. To
do so, we denote by $\overline{\omega}_{r},\overline{\omega}_{s},\overline{\omega}_{p}$
and $\overline{\omega}_{s}^{*}$ the numerical values of the cut-offs
calculated by the Bloch-Floquet analysis. Moreover, we also denote
by $\overline{a}_{\textrm{L}}$ and $\overline{a}_{\textrm{T}}$ the
numerical values of the slopes of the tangents to the acoustic curves
obtained via the Bloch-Floquet analysis. Note again that the third
(out-of-plane) acoustic branch is not present in this case since we
implemented a Bloch-Floquet analysis of a fully 2D metamaterial. In
Fig.\,\ref{fig:Dispersion-curves-of} we identify such numerical
quantities. 
The numerical values of the cut-off frequencies corresponding to Fig.\,\ref{fig:Dispersion-curves-of}
are given in Table \ref{table:cut-offs}. 
\begin{table}[H]
\begin{centering}
\begin{tabular}{cccc}
$\overline{\omega}_{r}$ & $\overline{\omega}_{s}$ & $\overline{\omega}_{s}^{*}$ & $\overline{\omega}_{p}$\tabularnewline[1mm]
\hline 
\noalign{\vskip1mm}
$\left[\mathrm{rad/s}\right]$ & $\left[\mathrm{rad/s}\right]$ & $\left[\mathrm{rad/s}\right]$ & $\left[\mathrm{rad/s}\right]$\tabularnewline[1mm]
\hline 
\hline 
\noalign{\vskip1mm}
$0.43\cdot10^{7}$ & $1.68\cdot10^{7}$ & $1.68\cdot10^{7}$ & $1.75\cdot10^{7}$\tabularnewline[1mm]
\end{tabular}
\par\end{centering}
\caption{\label{table:cut-offs}Numerical values of the cut-offs for the considered
metamaterial.}
\end{table}
Replacing in the last three equations (\ref{cut-offs})
the values of the cut-off frequencies given in Table \ref{table:cut-offs}
as well as the values of the elastic parameters given in Table\,\ref{table:micro inertae}b
and \ref{C_e}, we can uniquely determine $\eta_{1},\eta_{1}^{*}$
and $\eta_{3}$, obtaining the values in Table\,\ref{table:micro inertae}a.
\begin{table}[H]
\begin{centering}
\begin{tabular}{ccc}
$\eta_{1}$ & $\eta_{3}$ & $\eta_{1}^{*}$\tabularnewline[1mm]
\hline 
\noalign{\vskip1mm}
$\left[\mathrm{kg/m}\right]$ & $\left[\mathrm{kg/m}\right]$ & $\left[\mathrm{kg/m}\right]$\tabularnewline[1mm]
\hline 
\hline 
\noalign{\vskip1mm}
$9.3\cdot10^{-5}$ & $0.97\cdot10^{-5}$ & $3.19\cdot10^{-5}$\tabularnewline[1mm]
&(a)&
\end{tabular}
\qquad
\begin{tabular}{ccc}
	$\lambda_{\textrm{micro}}$ & $\mu_{\textrm{micro}}$ & $\mu_{\textrm{micro}}^{\ast}$\tabularnewline[1mm]
	\hline 
	\noalign{\vskip1mm}
	$\left[\textrm{GPa}\right]$ & $\left[\textrm{GPa}\right]$ & $\left[\textrm{GPa}\right]$\tabularnewline[1mm]
	\hline 
	\hline 
	\noalign{\vskip1mm}
	$5.27$ & $8.93$ & $8.33$\tabularnewline[1mm]
		&(b)&
\end{tabular}
\par\end{centering}
\caption{\label{table:micro inertae}Numerical values for (a) micro inertia parameters and (b) the $\cm$ parameters.}
\end{table}
The parameters of the relaxed micromorphic model which remain free
after these considerations are $\eta_{2},\overline{\eta}_{1},\overline{\eta}_{2},\overline{\eta}_{3},\overline{\eta}_{1}^{*}$.
To complete the fitting procedure, we start slowly increasing these
free parameters, starting from zero, so as to optimally fit dispersion
curves of Fig.\,\ref{fig:Dispersion-curves-of} both for $\vartheta=0$
and $\vartheta=\pi/4$. The order with which the free inertia parameters
are increased is related to the effect that such parameters have on
the dispersion curves. More particularly:
\begin{itemize}
\item $\eta_{2}$ and $\overline{\eta}_{2}$ have an effect on the acoustic
curves $\mathrm{LA}$ and $\mathrm{TA}$ and are adjusted to best
fit such curves,
\item the remaining parameters $\overline{\eta}_{1},\overline{\eta}_{1}^{*},\overline{\eta}_{3}$
are eventually increased only for fine-tuning the fitting. Their effect
is mainly visible for higher wavenumber (smaller wavelength).
\end{itemize}
As for the characteristic length $L_{c}$, we set here $L_{c}=0$.
Nevertheless, we know that a non-vanishing $L_{c}$ is a crucial point
for a finer fitting of the dispersion curves. On the other hand, this
task is really delicate and we need to postpone it to a further work
where a micro-inertia related to $\textrm{Curl }P_{,t}$ will also
be introduced.

Summarizing the results obtained up to now, we show in Table \ref{Numerical values}
the values for the inertiae and in Table \ref{Numerical values2}
a summary of all the elastic parameters computed before.

\begin{table}[H]
\begin{centering}
\begin{tabular}{ccccccccc}
$\rho$ & $\eta_{1}$ & $\eta_{2}$ & $\eta_{3}$ & $\eta_{1}^{*}$ & $\overline{\eta}_{1}$ & $\overline{\eta}_{2}$ & $\overline{\eta}_{3}$ & $\overline{\eta}_{1}^{*}$\tabularnewline[1mm]
\hline 
\noalign{\vskip1mm}
$\left[\mathrm{kg/m}^{3}\right]$ & $\left[\mathrm{kg/m}\right]$ & $\left[\mathrm{kg/m}\right]$ & $\left[\mathrm{kg/m}\right]$ & $\left[\mathrm{kg/m}\right]$ & $\left[\mathrm{kg/m}\right]$ & $\left[\mathrm{kg/m}\right]$ & $\left[\mathrm{kg/m}\right]$ & $\left[\mathrm{kg/m}\right]$\tabularnewline[1mm]
\hline 
\hline 
\noalign{\vskip1mm}
$1485$ & $9.3\cdot10^{-5}$ & $1\cdot10^{-7}$ & $0.97\cdot10^{-5}$ & $3.19\cdot10^{-5}$ & $4.8\cdot10^{-5}$ & $0$ & $0$ & $0$\tabularnewline[1mm]
\end{tabular}
\par\end{centering}
\caption{\label{Numerical values}Summary of the numerical values for inertia
parameters.}
\end{table}

\begin{table}[H]
\begin{centering}
\begin{tabular}{ccccccc}
\begin{tabular}{ccc}
$\le$ & $\mu_{e}$ & $\me^{*}$\tabularnewline[1mm]
\hline 
\noalign{\vskip1mm}
$\left[\textrm{GPa}\right]$ & $\left[\textrm{GPa}\right]$ & $\left[\textrm{GPa}\right]$\tabularnewline[1mm]
\hline 
\hline 
\noalign{\vskip1mm}
$-\,0.77$ & $17.34$ & $0.67$\tabularnewline[1mm]
\end{tabular} &  & %
\begin{tabular}{ccc}
$\lh$ & $\mh$ & $\mh^{*}$\tabularnewline[1mm]
\hline 
\noalign{\vskip1mm}
$\left[\textrm{GPa}\right]$ & $\left[\textrm{GPa}\right]$ & $\left[\textrm{GPa}\right]$\tabularnewline[1mm]
\hline 
\hline 
\noalign{\vskip1mm}
$5.27$ & $8.93$ & $8.33$\tabularnewline[1mm]
\end{tabular} &  & %
\begin{tabular}{c}
$\mc$\tabularnewline[1mm]
\hline 
\noalign{\vskip1mm}
$\left[\textrm{GPa}\right]$\tabularnewline[1mm]
\hline 
\hline 
\noalign{\vskip1mm}
$1.85\cdot10^{-3}$\tabularnewline[1mm]
\end{tabular} &  & %
\begin{tabular}{ccc}
$\lam$ & $\mum$ & $\mum^{*}$\tabularnewline[1mm]
\hline 
\noalign{\vskip1mm}
$\left[\textrm{GPa}\right]$ & $\left[\textrm{GPa}\right]$ & $\left[\textrm{GPa}\right]$\tabularnewline[1mm]
\hline 
\hline 
\noalign{\vskip1mm}
$1.74$ & $5.89$ & $0.62$\tabularnewline[1mm]
\end{tabular}\tabularnewline
\end{tabular}
\par\end{centering}
\caption{\label{Numerical values2}Summary of the numerical values for the
elastic parameters of the tetragonal relaxed micromorphic model in
2D. The macroscopic parameters of the resulting homogenized metamaterial
are also provided in the last Table. The values related to $\protect\ce$
still define a positive definite tensor.}
\end{table}

\begin{figure}[H]
\begin{centering}
\begin{tabular}{cc}
\includegraphics[scale=0.5]{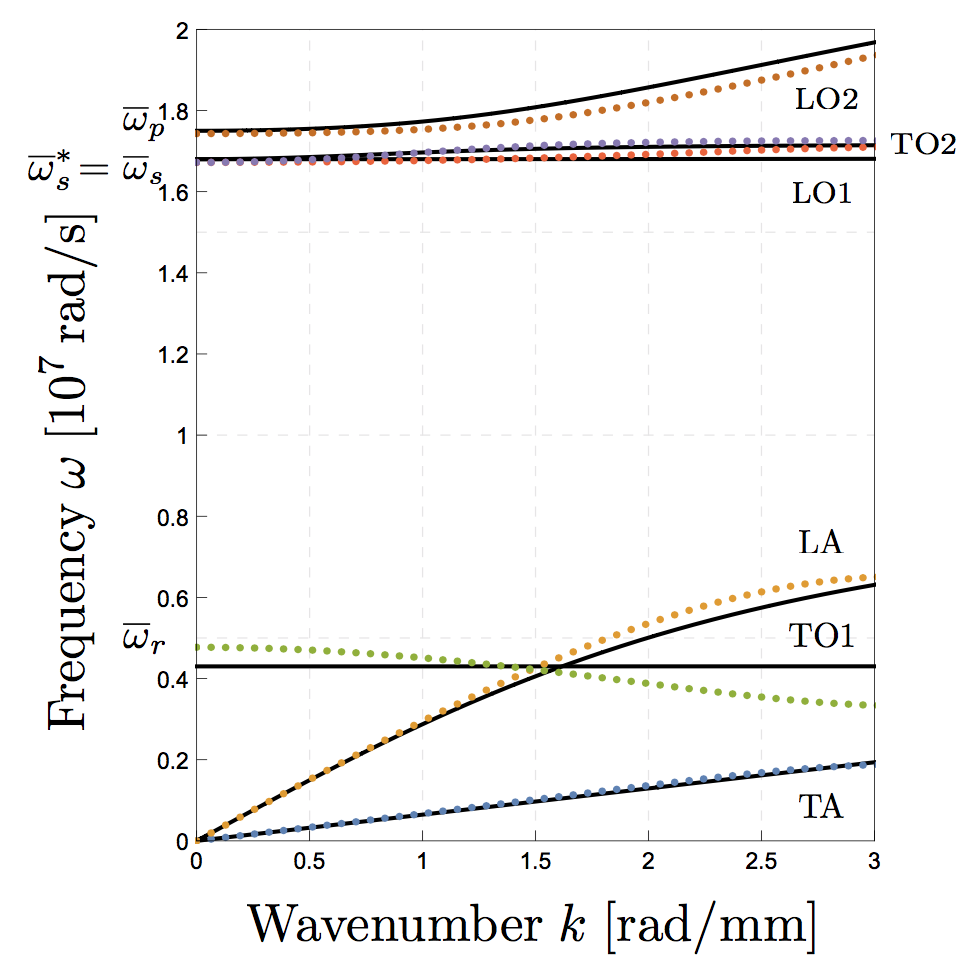} & \includegraphics[scale=0.5]{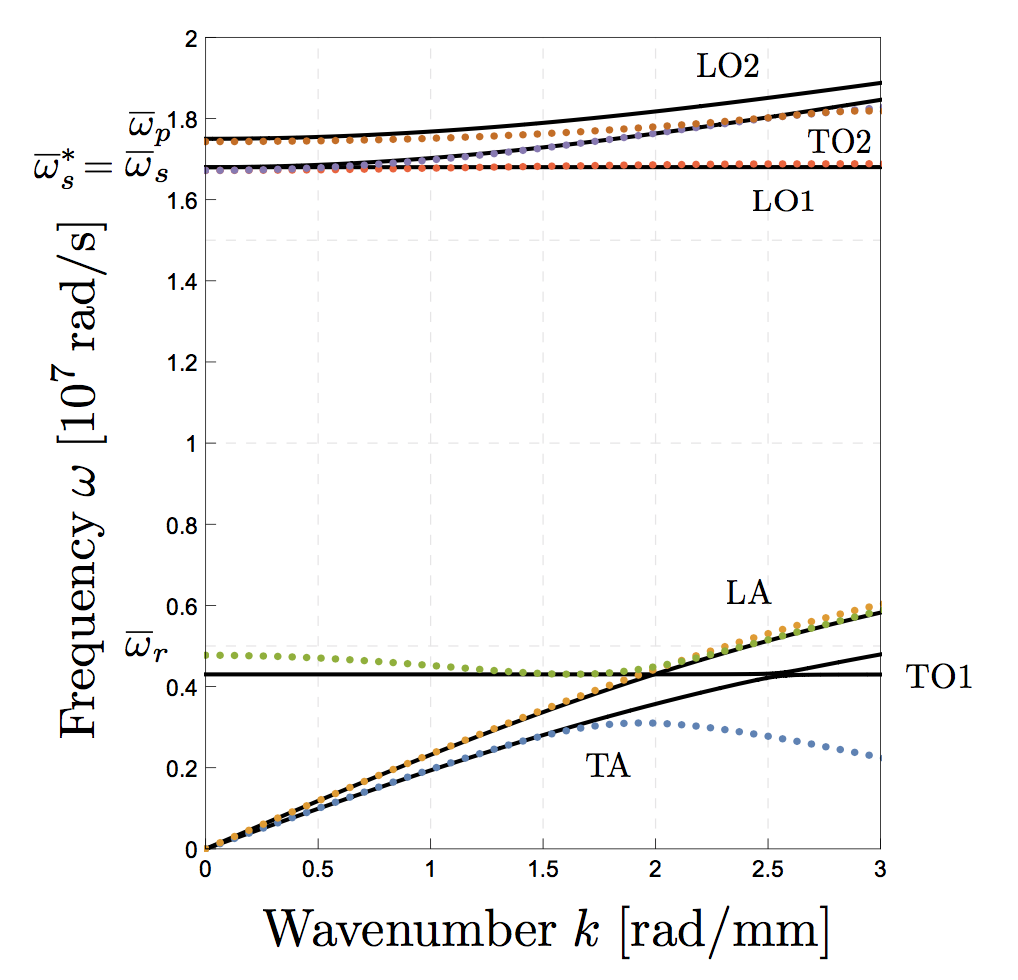}\tabularnewline
(a) & (b)\tabularnewline
\end{tabular}
\par\end{centering}
\caption{\label{fig:Comparison-between-our}Comparison between our relaxed
micromorphic continuum model and the COMSOL$^{\circledR}$ one. In
(a) we plot the dispersion branches for $\widehat{\boldsymbol{k}}=\left(1,0,0\right)$
and in (b) for $\widehat{\boldsymbol{k}}=\left(\sqrt{2}/2,\sqrt{2}/2,0\right)$.
Dotted lines represent COMSOL$^{\circledR}$ dispersion curves, continuous
lines represent the dispersion curves obtained with the relaxed micromorphic
model. The two directions $\widehat{\boldsymbol{k}}$ are used in
the fitting procedure.}
\end{figure}

In Fig.\,\ref{fig:Comparison-between-our} we show the obtained fitting
for $\vartheta=0$ and $\vartheta=\pi/4$. 

The result is quite satisfactory for a wide range of wavelengths.
The only relevant differences can be found in the curve $\mathrm{TO}_{1}$
both for $\vartheta=0$ and $\vartheta=\pi/4$ and $\mathrm{TA}$
at $\vartheta=\pi/4$. This discrepancy is mainly due to the fact
that at present the relaxed micromorphic model is not yet able to
give rise to decreasing dispersion curves. Such possibility will be
taken into account in future work by adding suitably non-local terms
in the kinetic energy and considering $L_{c}>0$.

The parameter calibration
that we present in Section 7  is the most natural and simple
approach imaginable using enriched continuum models. Indeed, the elastic
parameters are calibrated on simple mechanical tests on both macro
and micro specimens of the considered metamaterial. On the other hand,
dynamical parameters (micro-inertiae) are determined by imposing simple
relations on the cut-off frequencies. After that, very few parameters
remain free and they are slowly varied to improve the fitting of the
dispersion curves at higher wavenumbers (small wavelengths). We claim that we are providing a transparent and efficient
characterization of the considered mechanical metamaterial by means
of an enriched continuum model.

Our method is physics-based and it is far-away from parameter fittings
that are often provided when dealing with the superposition of generalized
continuum models to phenomenological data: we do not need to calibrate
a huge number of parameters by using ``ad hoc'' optimization methods,
we just obtain the fitting as a simple consequence of our mechanical
observations.

We also remark that slight differences can be found in the fitting
for some of the higher optic curves when considering high wavenumbers
(wavelength smaller than twice the unit-cell), while in general most
of the dispersion curves fit well the overall behavior also for wavelength
which go down to the size of the unit-cell. We will show in further
works that generalizing the expression of the kinetic energy will
allow to obtain an even better fitting for the whole considered range
of wavelengths.

\section{Prediction of dispersion and anisotropy in tetragonal metamaterials }

In this Section we show the capability of the anisotropic relaxed
micromorphic model to describe complex phenomena in specific metamaterials.
We have already seen in Section \ref{sec:metamaterial} that the metamaterial
targeted in this paper has a tetragonal symmetry. Moreover, in Section \ref{sec:Fitting} we have developed
the procedure allowing to calibrate the material parameters of the
relaxed micromorphic model on the considered metamaterial.

In the present Section, we show that the relaxed micromorphic model,
as calibrated on the considered tetragonal metamaterial, is able to
simultaneously reproduce very complex observable macroscopic phenomena,
namely
\begin{itemize}
\item the dispersive behavior both for the acoustic and the optic curves,
\item anisotropic (tetragonal) mechanical behavior of the considered metamaterial
for the first six modes.
\end{itemize}
The first characteristic, i.e., the dispersive behavior of the metamaterial,
has already been underlined in the previous Section when noticing
that the dispersion curves are not straight lines, but curves. This
means that the speed of propagation of each mode is not a constant
(as is the case in classical Cauchy continua) but varies when changing
the wavelength of the travelling wave. We show again in this Section
how the dispersive behavior of the considered metamaterial can be
highlighted by introducing the concept of phase velocity. The \textbf{phase
velocity} is defined as the ratio $\omega\left(k\right)/k$ and, in
dispersive media, changes when changing the wavenumber (or equivalently
the wavelength).

The phase velocity also changes when changing the direction of propagation
of the travelling wave if the considered medium is not isotropic.
Both such features of the phase velocity are easily understandable
since:
\begin{itemize}
\item the speed of propagation of waves reasonably changes when the travelling
wave reaches a wavelength which is comparable to the characteristic
size of the underlying microstructure. Such wavelength are easily
attainable for the most common metamaterials for which the microstructure
has typically the size of millimeters or more,
\item if the medium is anisotropic, i.e., if its deformation response varies
when varying the direction of application of the external load, it
is clear that the speed of propagation of waves inside the medium
changes as well when changing the direction of propagation of the
travelling wave itself.
\end{itemize}
We can simultaneously show both such characteristics (anisotropy and
dispersion) by looking at the polar plots of the phase velocity $\omega\left(k\right)/k$
for each mode and for different values of the wavenumber $k$ and
of the angle $\vartheta$ giving the direction of propagation of the
travelling wave. 

\begin{figure}[h]
\begin{centering}
\begin{tabular}{c}
\textbf{\large{}limit macroscopic Cauchy model}\tabularnewline
\end{tabular}
\par\end{centering}
\begin{centering}
\begin{tabular}{cc}
 & \tabularnewline
\includegraphics[scale=0.5]{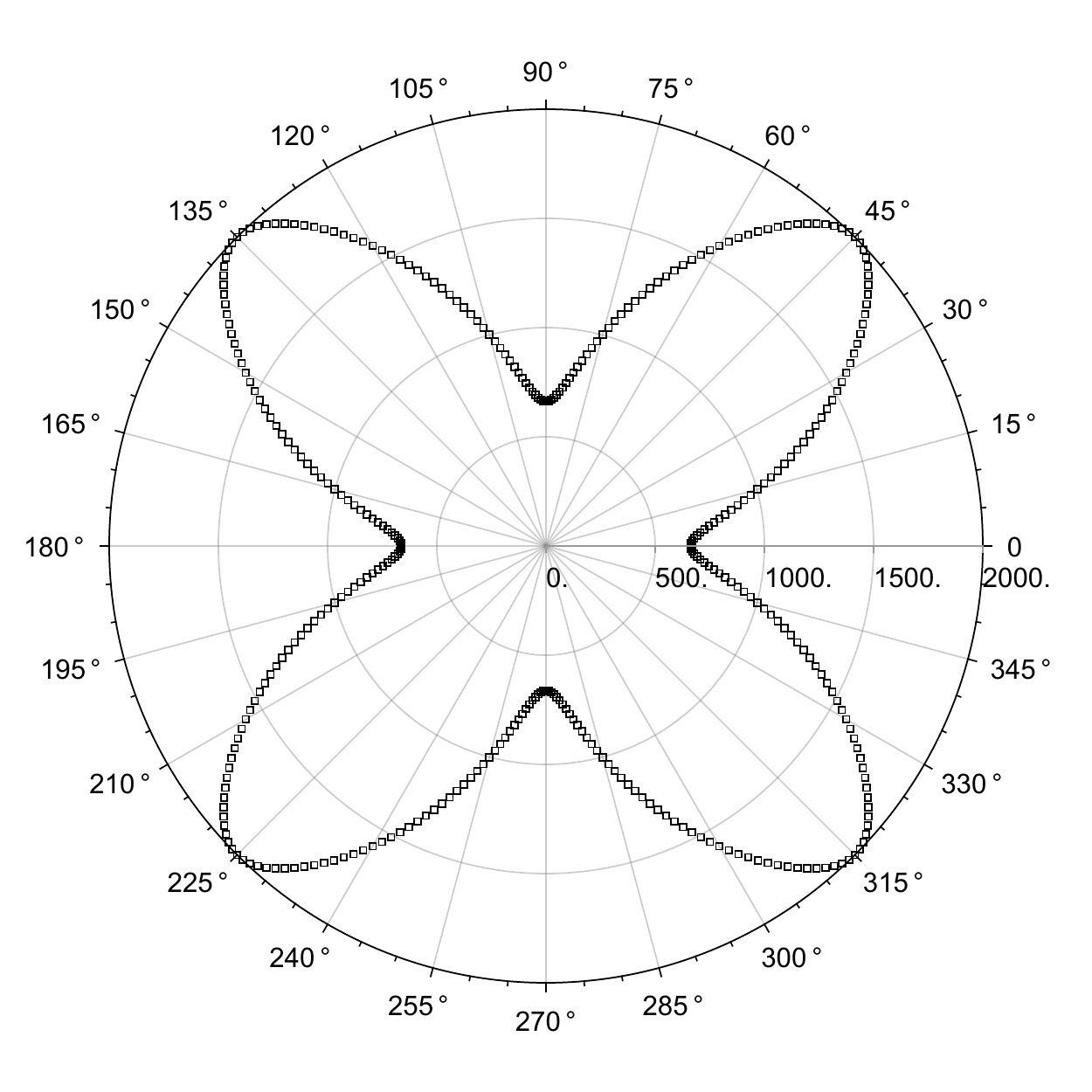} & \includegraphics[scale=0.5]{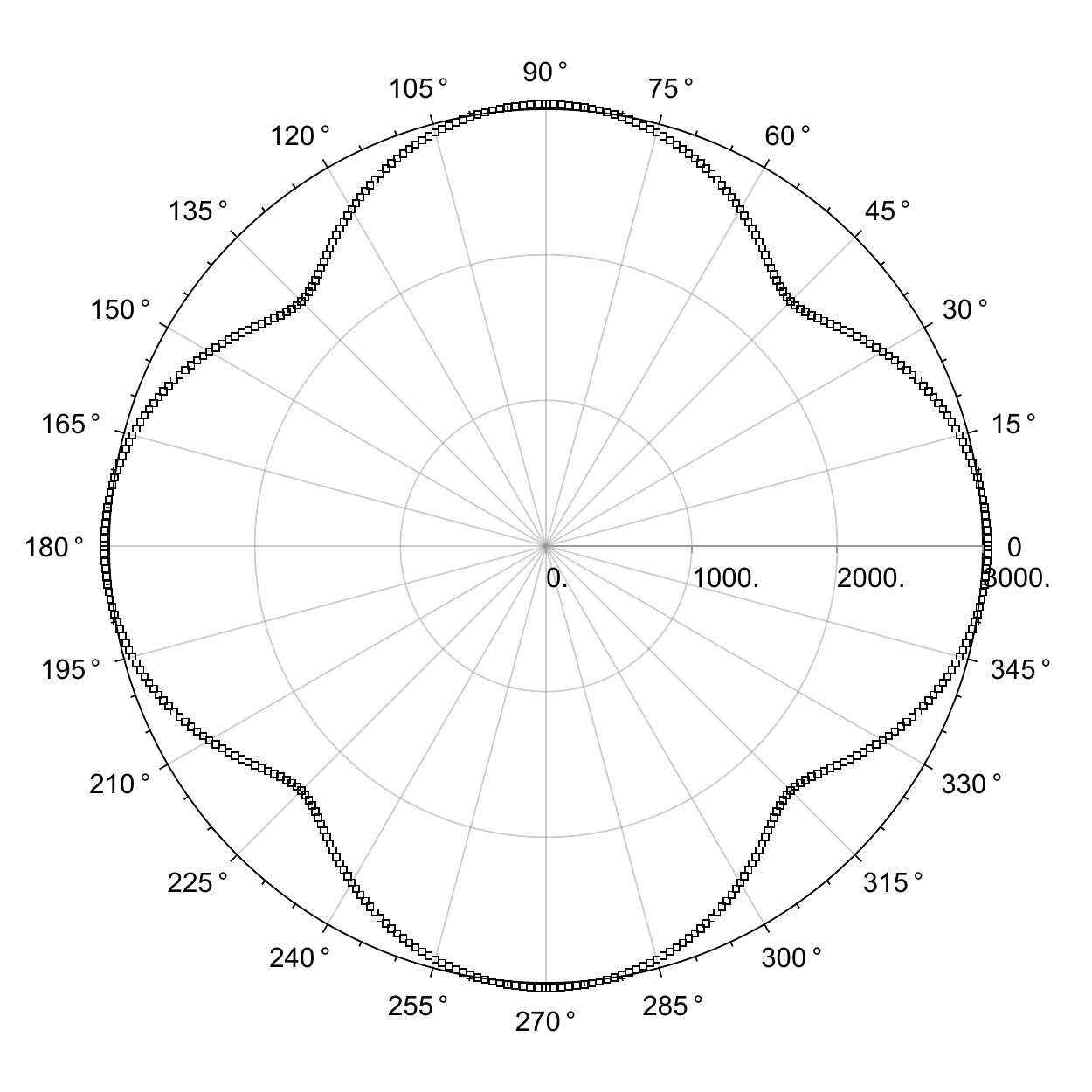}\tabularnewline
\textbf{TA} & \textbf{LA}\tabularnewline
\end{tabular}
\par\end{centering}
\caption{\label{fig:Phase-velocity-as1-1} Polar plots of the phase velocity
$\omega/k$ for the two acoustic modes of the equivalent tetragonal
Cauchy medium, the elastic parameters of which are related to those
of the relaxed micromorphic model through formulas (\ref{eq:mue-muh})
and using the values given in Table \ref{Numerical values2}. In each
plot, each radial line from zero indicates the direction of propagation
of the travelling wave and the length of any segment going from zero
to a point of the curve is the measure of the phase velocity $\omega/k$
in the considered direction.}
\end{figure}

We start presenting the case of a classical tetragonal non-dispersive
Cauchy medium (see Fig.\,\ref{fig:Phase-velocity-as1-1}).

In Fig.\,\ref{fig:Phase-velocity-as1-1} we show the polar plots
of the phase velocity $\omega/k$ for the tetragonal Cauchy continuum
which is obtained as a limiting case of the relaxed micromorphic model
as fitted on the considered tetragonal metamaterial. More particularly,
Fig. \ref{fig:Phase-velocity-as1-1} shows that:
\begin{enumerate}
\item the two acoustic modes which are described by the Cauchy theory do
not describe dispersion. In fact, since in Cauchy theory the dispersion
curve is a straight line, once fixed the value of $k$, the phase
velocity $\omega/k$ takes a constant value for any chosen value of
$k$. It is for this reason that, in Cauchy media, we find only one
curve in the polar plot of the phase velocity.
\item Cauchy theory can only describe the first two acoustic modes without
any possibility of describing higher modes.
\end{enumerate}
According to these observations, it becomes clear that a Cauchy theory
is too restrictive for describing the rich behavior of metamaterials. 

As already remarked, second gradient theories could somewhat improve
the description of the dispersive behaviors with respect to the Cauchy
theory, but in any case only the first 2 modes could be analyzed in
the 2D case.

\begin{figure}[H]
	\begin{centering}
		\begin{tabular}{ccc}
			& \textbf{\large{}relaxed micromorphic model} & \textbf{\large{}Bloch-Floquet analysis}\tabularnewline
			&  & \tabularnewline
			\textbf{\LARGE{}}%
			\begin{tabular}{c}
				\noalign{\vskip-6.5cm}
				\textbf{\LARGE{}TA}\tabularnewline
			\end{tabular} & \includegraphics[scale=0.5]{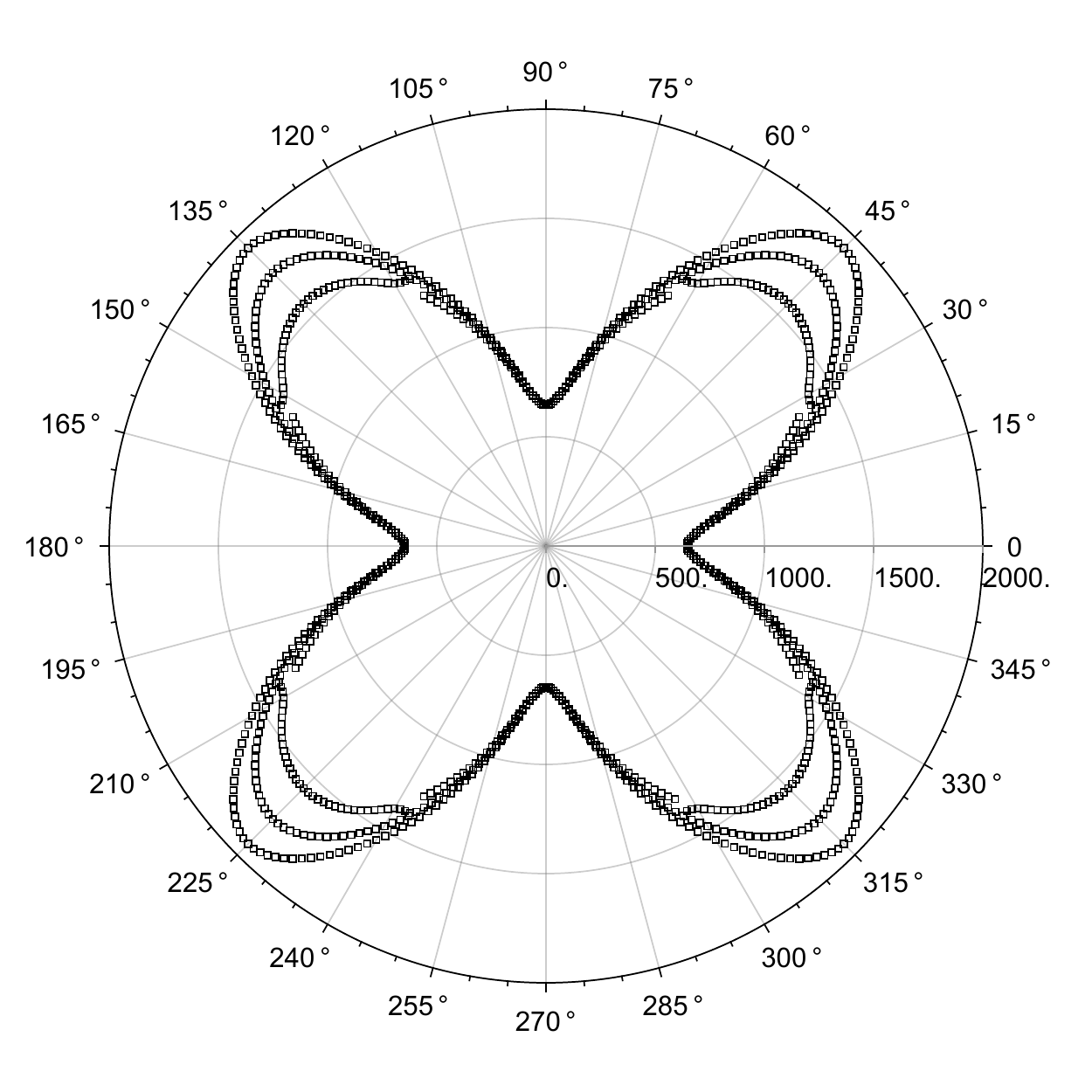} & \includegraphics[scale=0.5]{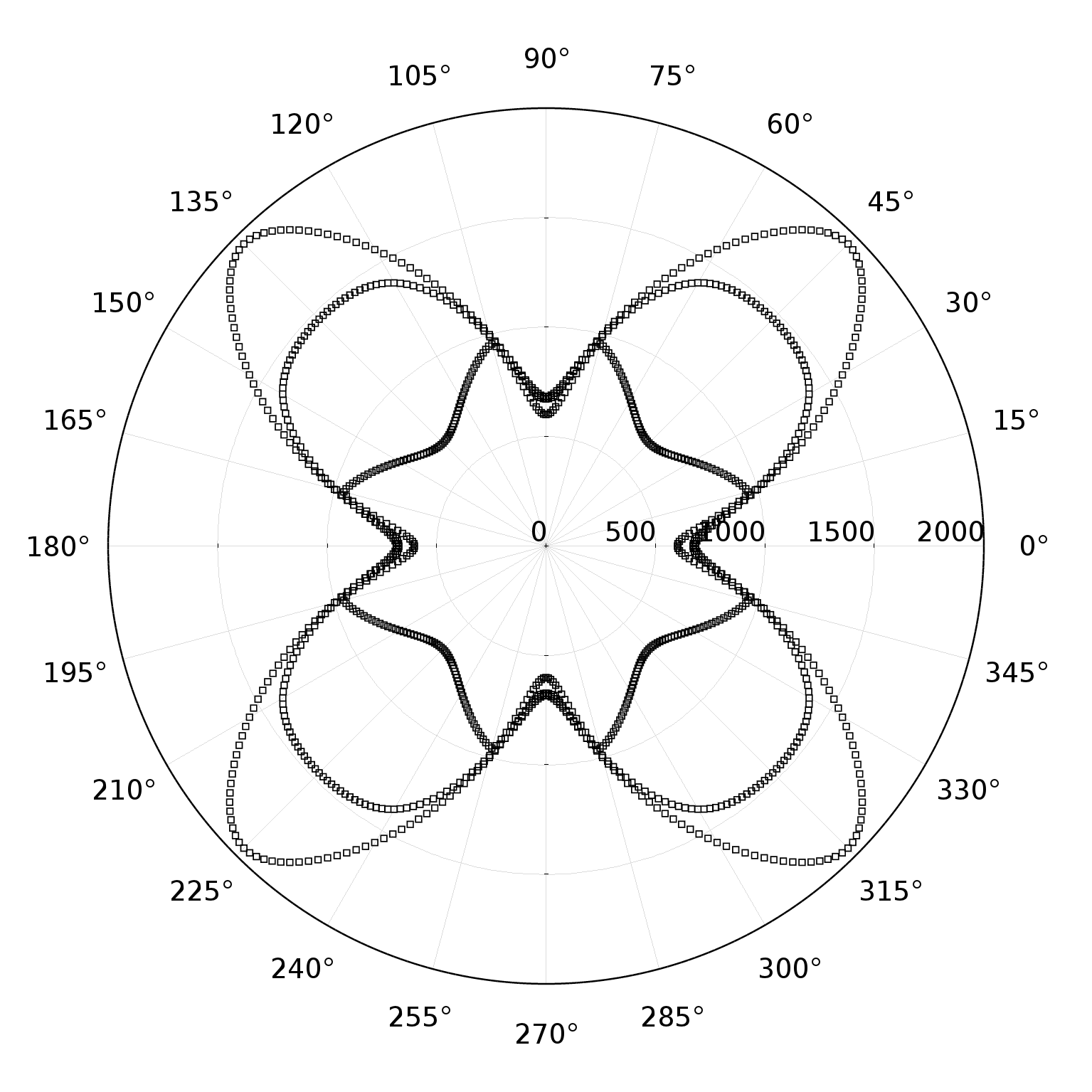}\tabularnewline
			\textbf{\LARGE{}}%
			\begin{tabular}{c}
				\noalign{\vskip-6.5cm}
				\textbf{\LARGE{}LA}\tabularnewline
			\end{tabular} & \includegraphics[scale=0.5]{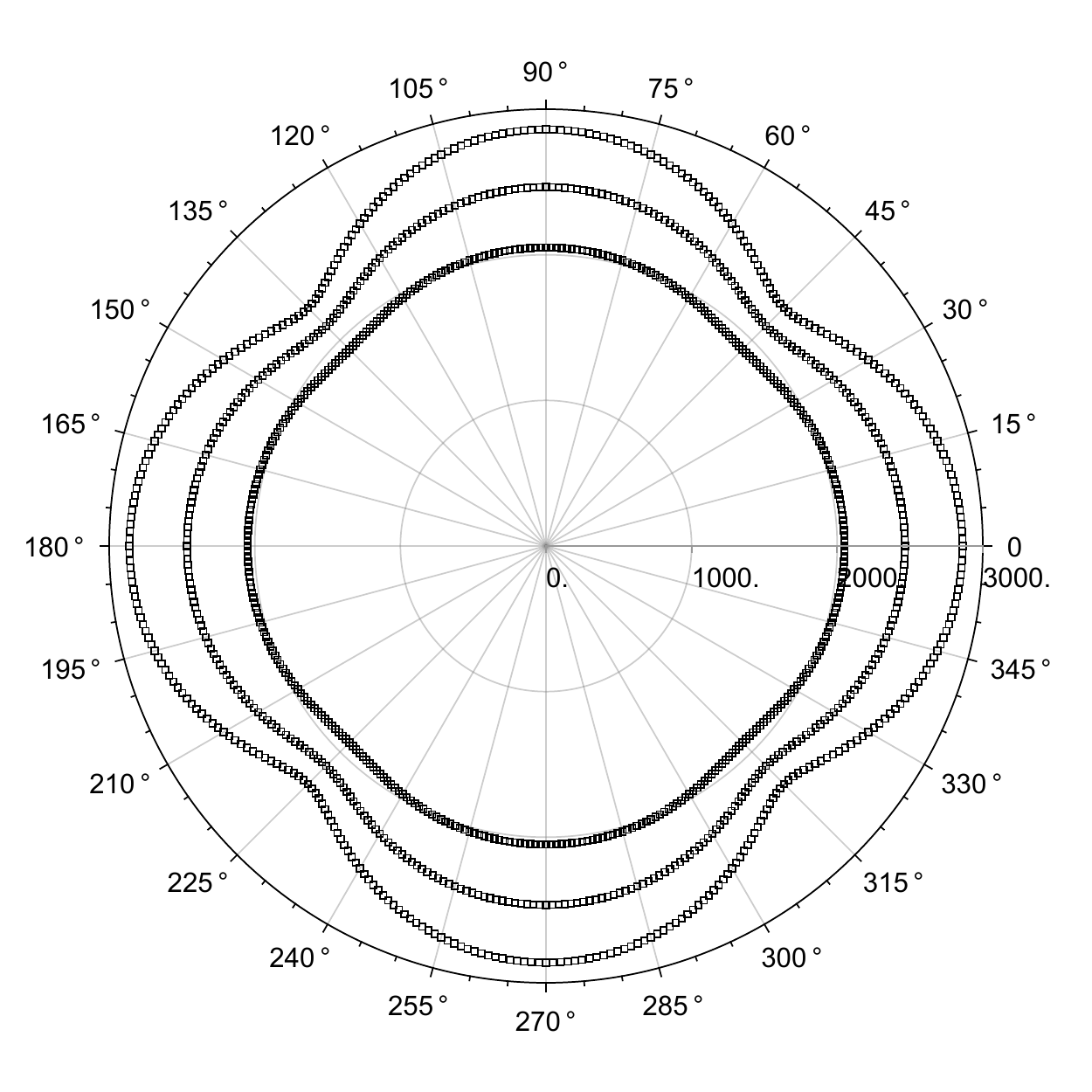} & \includegraphics[scale=0.5]{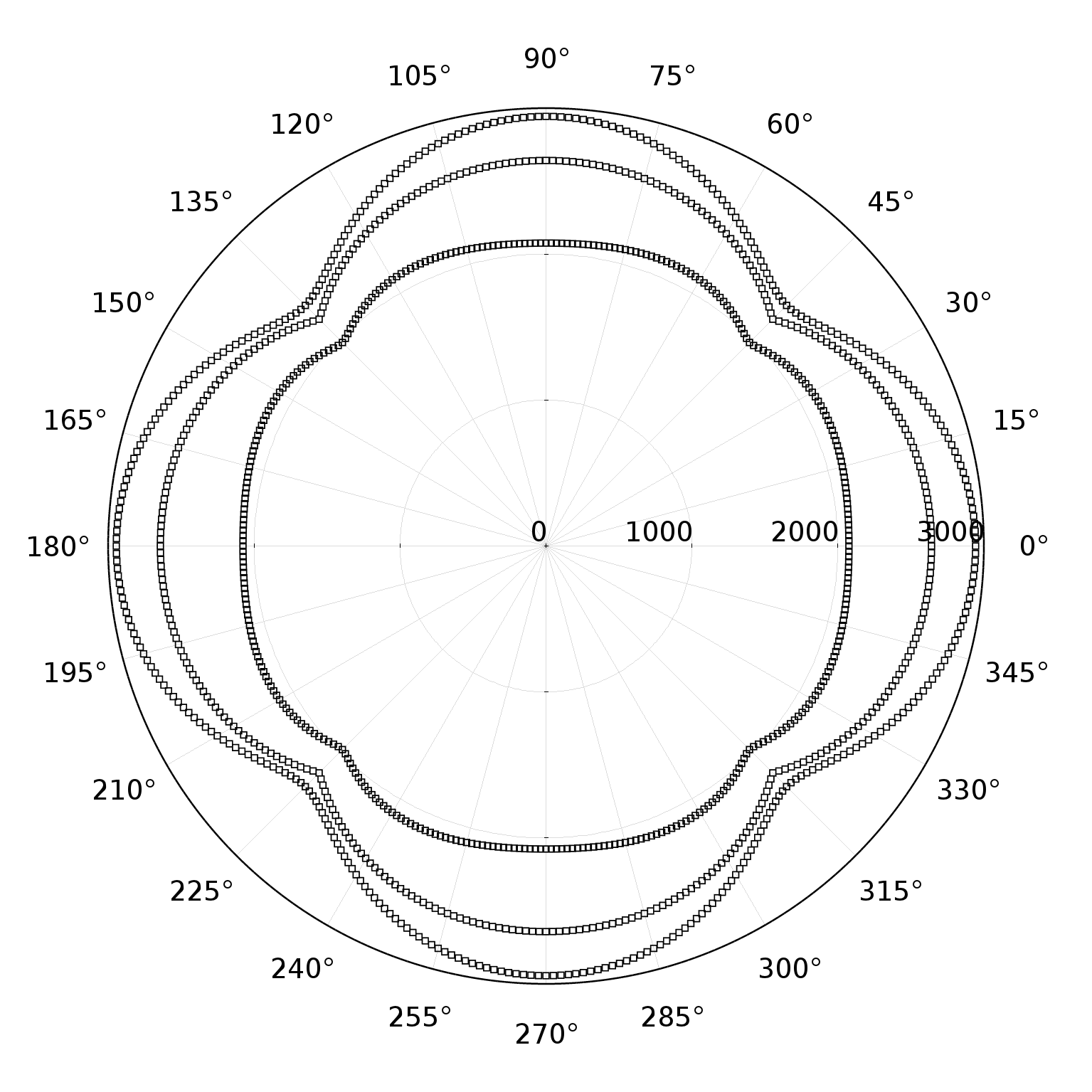}\tabularnewline
			\textbf{\LARGE{}}%
			\begin{tabular}{c}
				\noalign{\vskip-6.5cm}
				\textbf{\LARGE{}TO1}\tabularnewline
			\end{tabular} & \includegraphics[scale=0.5]{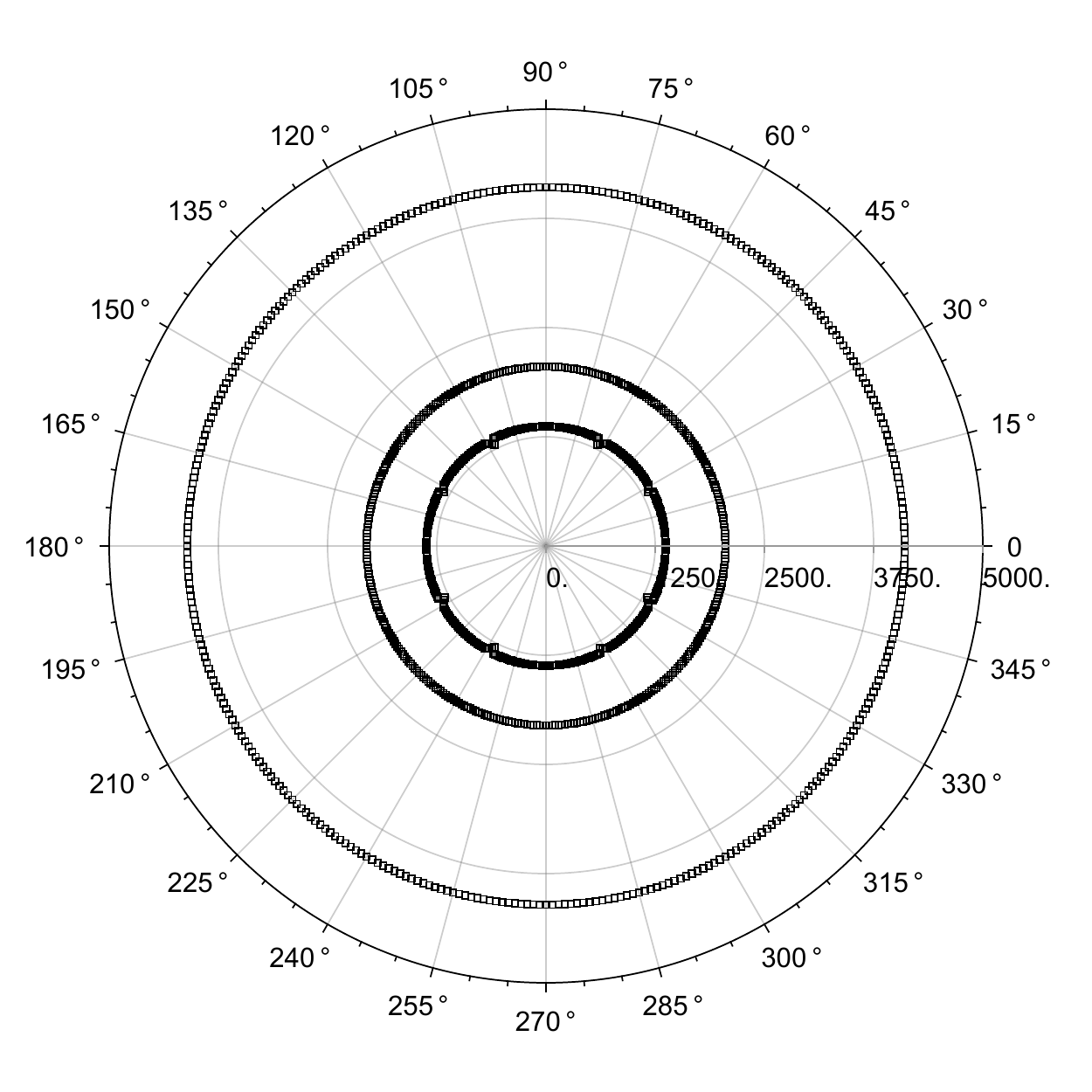} & \includegraphics[scale=0.5]{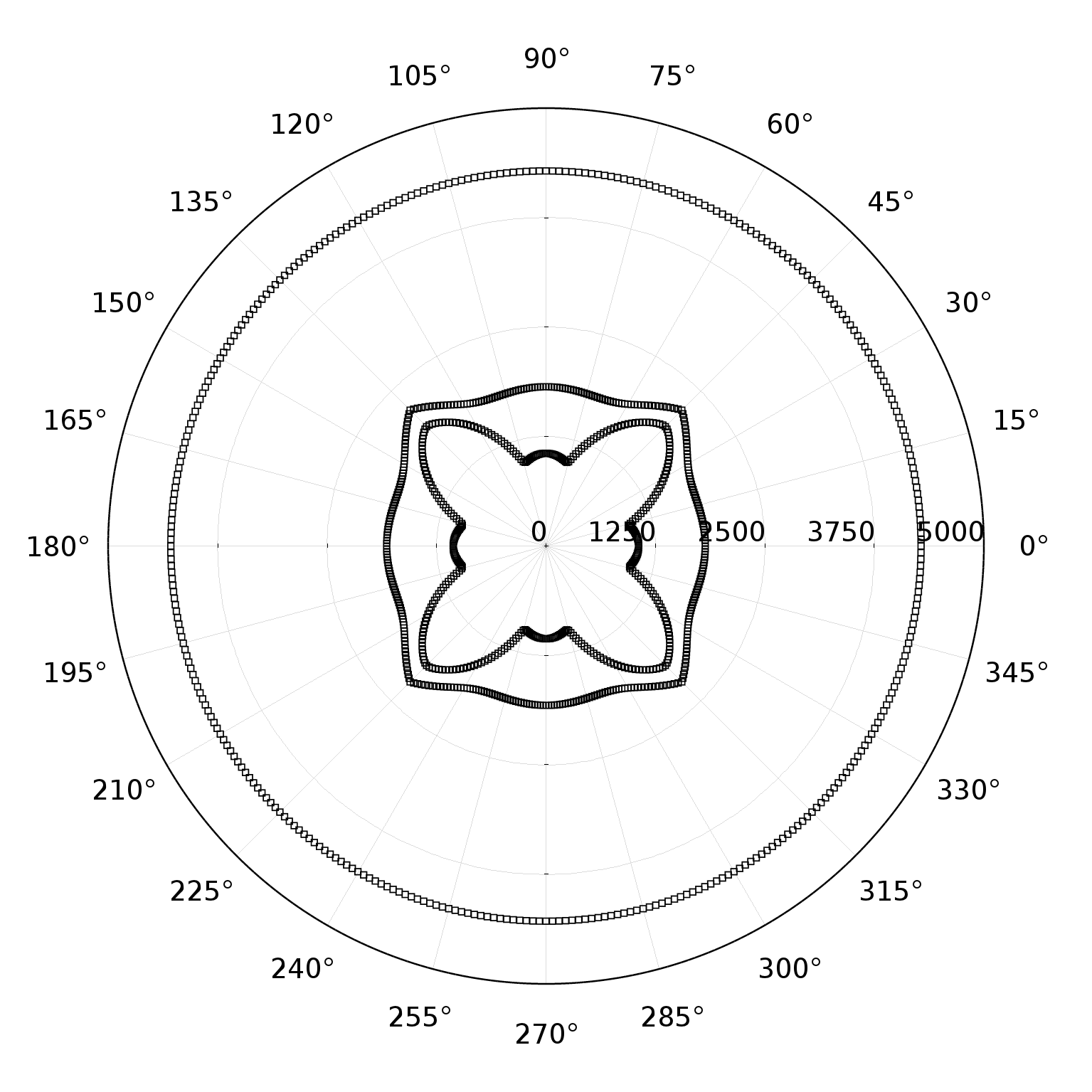}\tabularnewline
		\end{tabular}
		\par\end{centering}
	\caption{\label{fig:Phase-velocity-as1}Phase velocity as a function of the
		direction of wave propagation $\widehat{\boldsymbol{k}}$ for the
		first three modes as calculated with the relaxed micromorphic model
		(left) and with the Bloch-Floquet analysis (right). The plotted curves
		have been calculated for the values of the wave number equal to ${\displaystyle \pi,\frac{2\pi}{3},\frac{\pi}{3}\:\frac{\left[\mathrm{rad}\right]}{\left[\mathrm{mm}\right]}}$
		. For any curve, the distance from the center of the circle to a point
		of the curve itself gives the value of the phase velocity $\omega/k$.
		More external curves are relative to lower values of $k$, while the
		curves become closer to the origin when increasing the value of $k$.
		The most internal curve corresponds to a wavelength comparable to
		the unit-cell.}
\end{figure}

As far as the relaxed micromorphic model is concerned, we will show
in the remainder of this Section that it is able to describe:
\begin{itemize}
\item not only the first two, but also other modes related to higher frequencies,
\item anisotropy, not only for the first two, but also for the other modes,
\item dispersion for all the considered modes (here the first 6).
\end{itemize}
We can start remarking from Fig.\,\ref{fig:Phase-velocity-as1} that
the transverse acoustic mode is perfectly described for lower values
of $k$ (external curve), while some differences with the Bloch-Floquet
analysis of the metamaterial arise for higher values of $k$ (more
internal curves).

The larger difference can be detected for waves travelling at $45\text{\textdegree}$.
This is consistent with Fig. \ref{fig:Comparison-between-our}(b) in
which it can be easily seen that the transverse wave $\textrm{TA}$
calculated via the relaxed micromorphic model starts diverging from
the one calculated via the Block-Floquet analysis when increasing
the value of $k$. This means that, as far as the transverse acoustic
mode $\textrm{TA}$ is concerned, its description via the relaxed
micromorphic model is less accurate when considering propagation at
$45\text{\textdegree}$ and higher wavenumbers (smaller wavelengths).
As a matter of fact, the description of the $\textrm{TA}$ mode at
$45\text{\textdegree}$ starts being less accurate for wavelengths
twice the characteristic size of the unit-cell and smaller.

This lack of accuracy for small wavelengths is not present for the
longitudinal acoustic mode $\textrm{LA}$ for which the relaxed micromorphic
model describes well the behavior of the $\textrm{LA}$ curve for
all directions of propagation and for wavelengths which become very
small and even comparable to the size of the unit-cell.

The behavior of the curve $\textrm{TO}_{1}$ has to be considered
carefully. Indeed, we fitted our relaxed micromorphic model to the
$\textrm{TO}_{1}$ dispersion curve with an almost horizontal curve.
We are hence able to recover for this mode the average dispersive
behavior, but not the true patterns of the phase velocity which show
anisotropic behaviors for higher wavenumbers (smaller wavelengths).
This behavior will be improved in further works when the non-locality
of the matematerial will be considered by considering $L_{c}>0$ and
by adding a term $\curl\,P_{,t}$ in the kinetic energy.

We also recall that the relaxed micromorphic model is able to catch
some other essential features of the description of the considered
metamaterial, such as the presence of band-gaps (see Fig.\ref{fig:Comparison-between-our}).

Based on some preliminary studies, we already know that the typical
decreasing behavior of the $\textrm{TO}_{1}$ curve obtained via the
Bloch-Floquet analysis can be described by the relaxed micromorphic
model if we allow it to include non local effects. In other words,
if we consider the case $L_{c}>0$, a better and finer fitting of
the $\textrm{TO}_{1}$ curve could be obtained. Nevertheless, the
study of this additional case deserves more attention since an extra
micro-inertia term involving $\textrm{Curl}\,P_{,t}$ should also
be considered. We will hence treat such complete non-local cases in
further works, building on the results presented here.

\begin{figure}[H]
	\begin{centering}
		\begin{tabular}{ccc}
			& \textbf{\large{}relaxed micromorphic model} & \textbf{\large{}Bloch-Floquet analysis}\tabularnewline
			&  & \tabularnewline
			\textbf{\LARGE{}}%
			\begin{tabular}{c}
				\noalign{\vskip-6.5cm}
				\textbf{\LARGE{}LO1}\tabularnewline
			\end{tabular} & \includegraphics[scale=0.5]{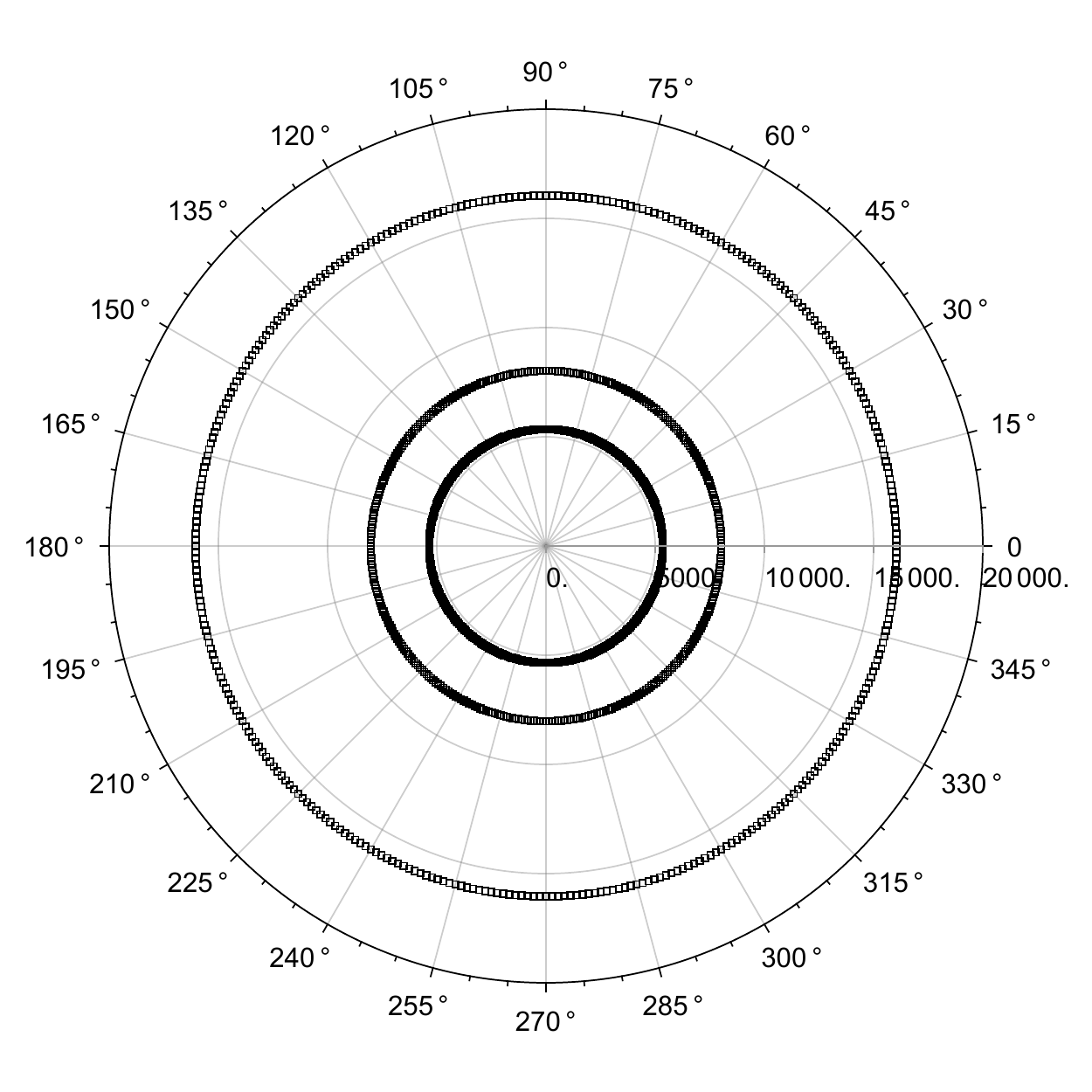} & \includegraphics[scale=0.5]{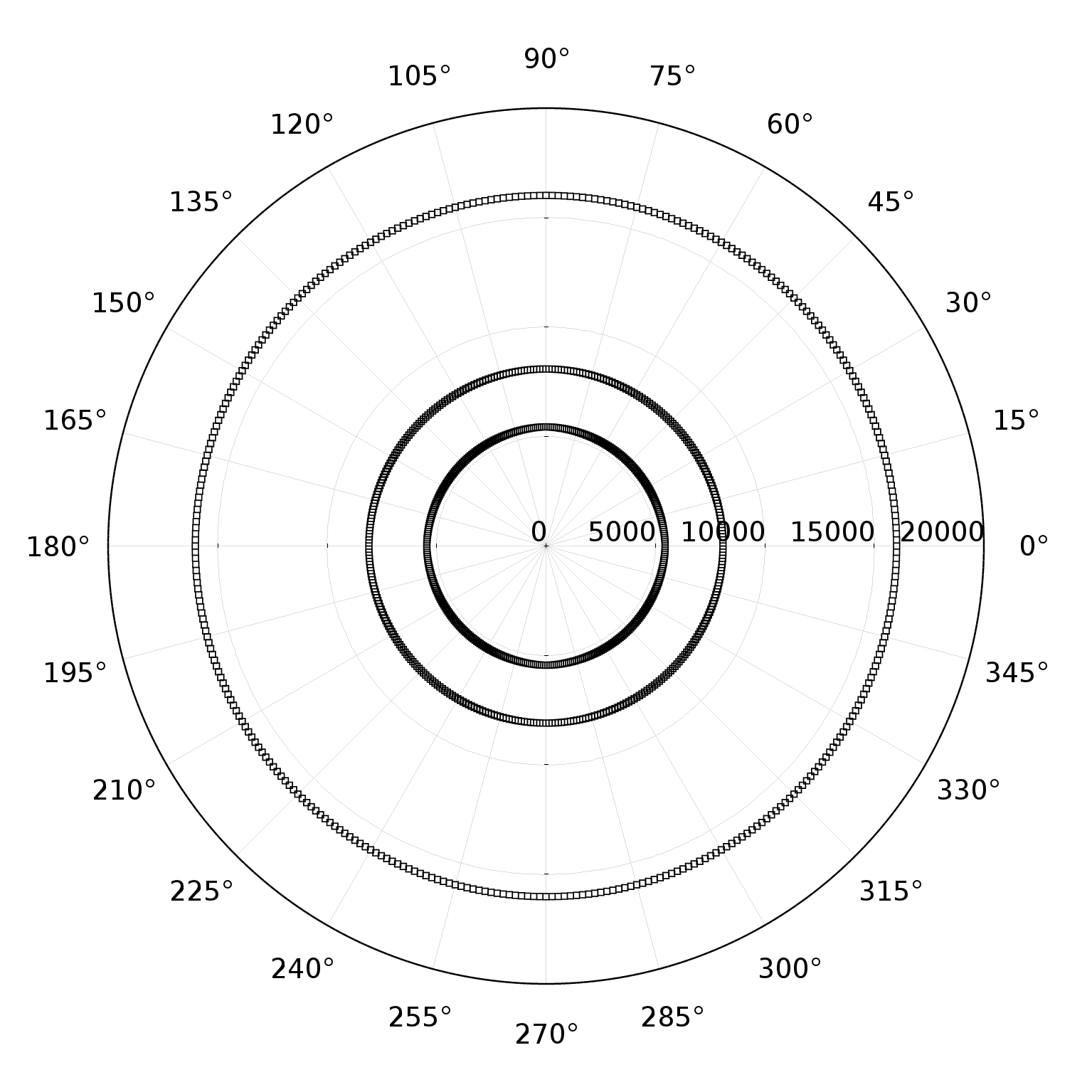}\tabularnewline
			\textbf{\LARGE{}}%
			\begin{tabular}{c}
				\noalign{\vskip-6.5cm}
				\textbf{\LARGE{}TO2}\tabularnewline
			\end{tabular} & \includegraphics[scale=0.5]{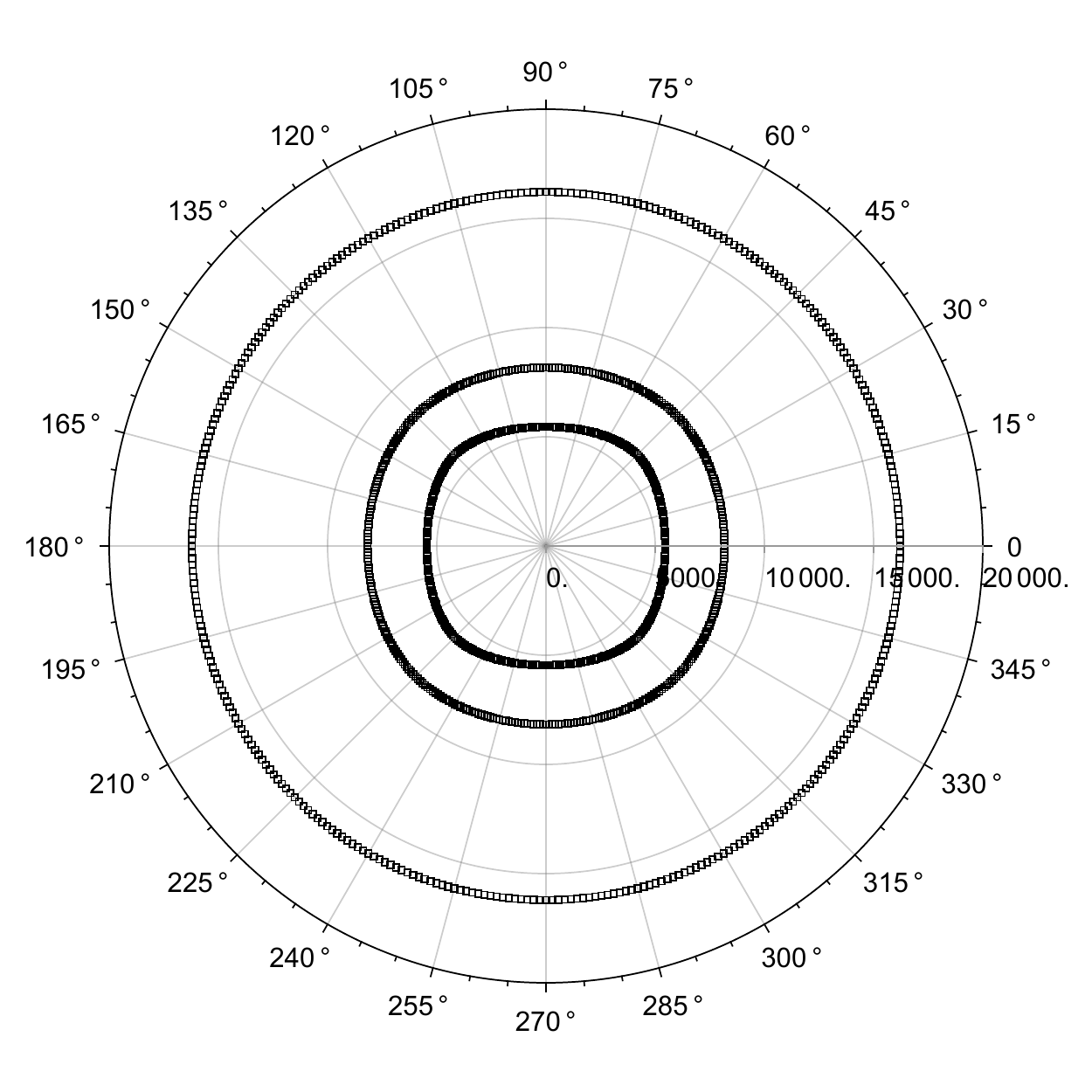} & \includegraphics[scale=0.5]{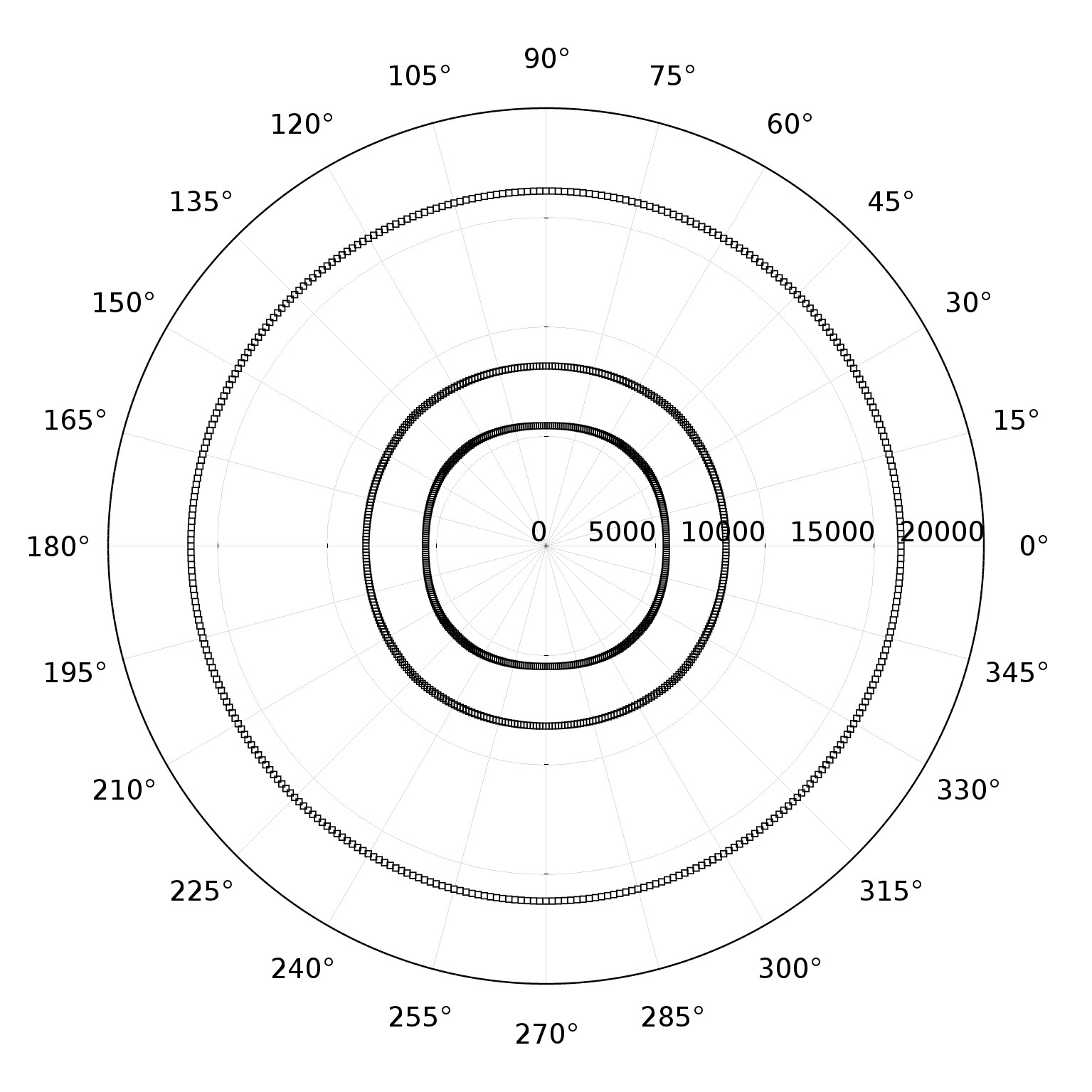}\tabularnewline
			\textbf{\LARGE{}}%
			\begin{tabular}{c}
				\noalign{\vskip-6.5cm}
				\textbf{\LARGE{}LO2}\tabularnewline
			\end{tabular} & \includegraphics[scale=0.5]{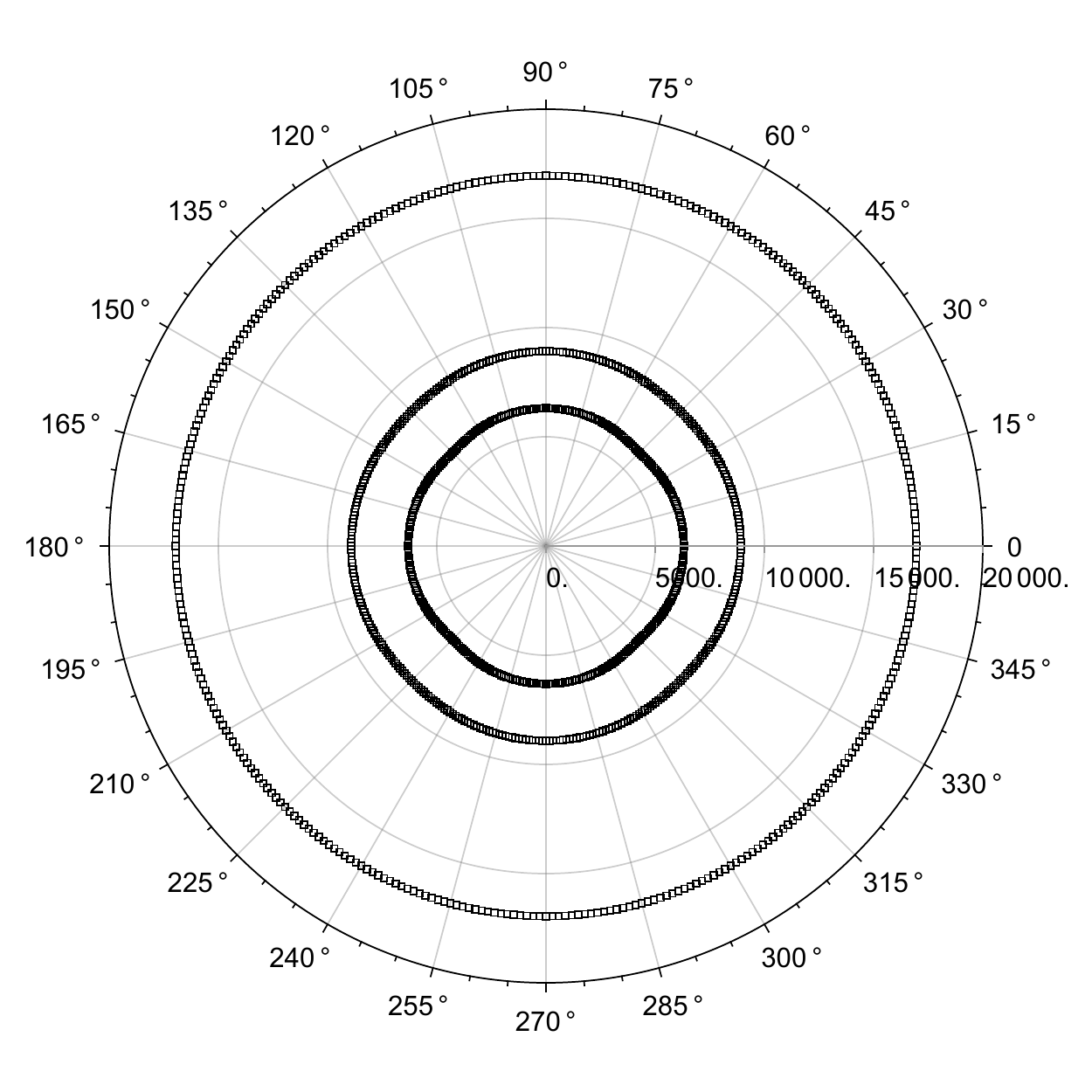} & \includegraphics[scale=0.5]{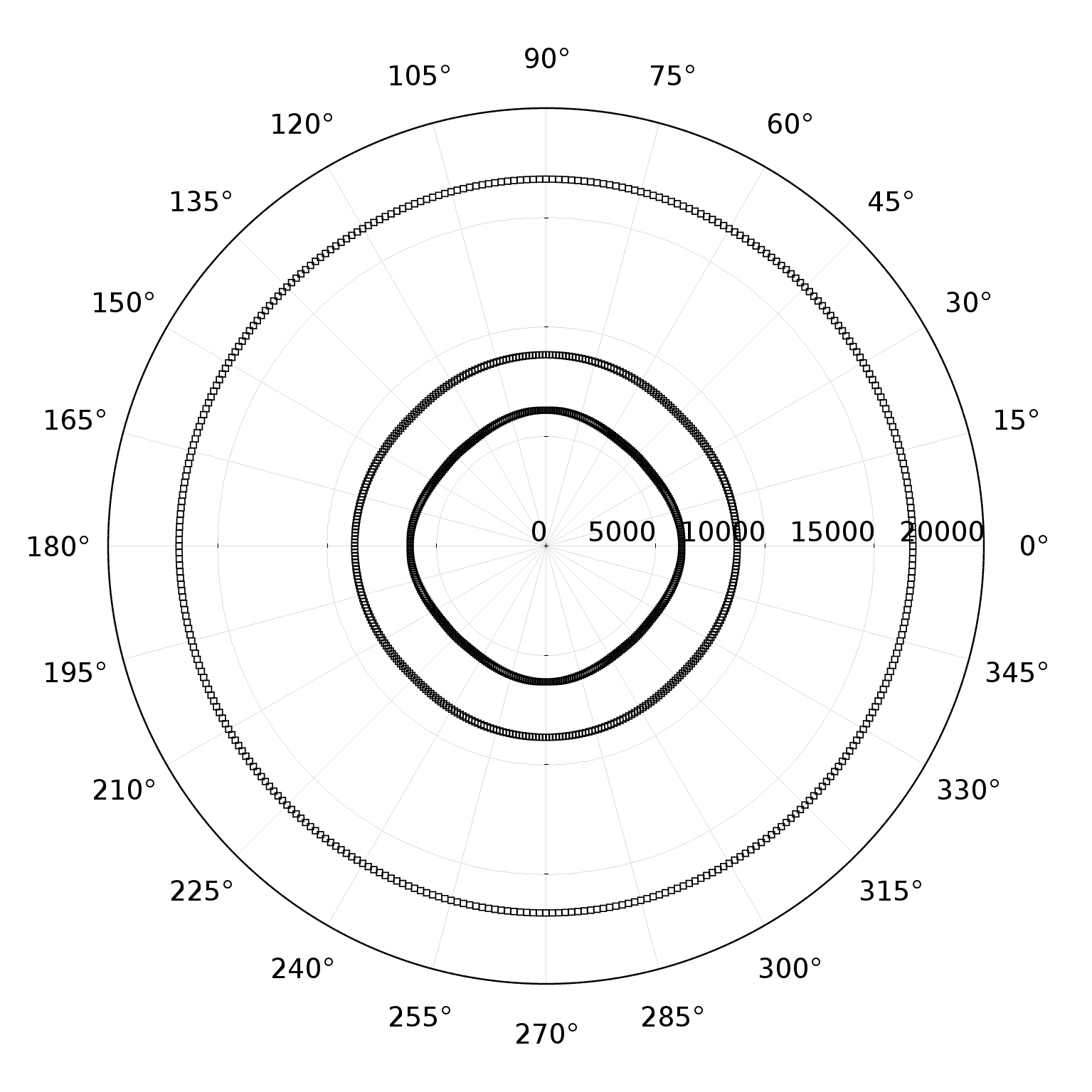}\tabularnewline
		\end{tabular}
		\par\end{centering}
	\caption{\label{fig:Phase-velocity-as2}Phase velocity as a function of the
		direction of wave propagation $\widehat{\boldsymbol{k}}$ for higher
		modes as calculated with the relaxed micromorphic model (left) and
		with the Bloch-Floquet analysis (right). The plotted curves have been
		calculated for the values of the wave number equal to ${\displaystyle \pi,\frac{2\pi}{3},\frac{\pi}{3}\:\frac{\left[\mathrm{rad}\right]}{\left[\mathrm{mm}\right]}}$
		. For any curve, the distance from the center of the circle to a point
		of the curve itself gives the value of the phase velocity $\omega/k$.
		More external curves are relative to lower values of $k$, while the
		curves become closer to the origin when increasing the value of $k$.
		The most internal curve corresponds to a wavelength comparable to
		the unit-cell.}
\end{figure}

In Fig. \ref{fig:Phase-velocity-as2} we can see that the relaxed
micromorphic model describes almost perfectly both dispersion and
anisotropy for the higher optic modes.

\section{Conclusion and further perspectives}

In this paper we have specialized the general anisotropic relaxed
micromorphic model developed in \cite{barbagallo2016transparent}
to the case of tetragonal symmetry. We show that this particular symmetry
class allows us to describe the anisotropic behavior of a band-gap
metamaterial with specific tetragonal microstructure. 

We explicitly show the true advantage of using the relaxed micromorphic
model \cite{barbagallo2016transparent} in describing the mechanical
behavior of metamaterials by introducing only standard fourth order
elasticity tensors (in Voigt-notation) as it is the case for classical
elasticity. This efficient theoretical framework allows us to avoid
unnecessary complexifications related, e.g., to the introduction of
elastic tensors of order higher than four, as it is the case for higher
gradient elasticity (see, e.g., \cite{auffray2013matrix,olive2014symmetry,olive2013symmetry}).
Indeed, the study of the anisotropy classes of the tensors appearing
in the relaxed micromorphic model follows the classical lines and
does not require any extra development.

Then we specialize the anisotropic relaxed micromorphic
model to the case of tetragonal symmetry. As a second step, we set up a fitting procedure
to determine the values of the parameters of the relaxed micromorphic
model by i) computing the purely elastic parameters via suitably conceived
static tests in \cite{d2019Identification} and ii) obtaining the values of dynamical parameters
by superimposing the dispersion curves obtained with the relaxed micromorphic
model to the corresponding ones obtained with a classical Bloch-Floquet
analysis. The relaxed micromorphic model is a ``macroscopic continuum''
homogenized model which is able to reproduce the response of the selected
metamaterial with only few material parameters which do not depend
on frequency.

The advantage of our continuum model can be found in the perspective
of modeling metamaterials in a simplified framework with transparent
mechanical interpretation and thus providing a concrete possibility
for the design of relatively complex metastructures by means of
 a relatively simplified model.

Moreover, the fact that our enriched continuum model is available
allows us to simplify the study of other problems that would be otherwise
difficult to treat, such as, e.g., the study of interfaces between
different anisotropic metamaterials.

The pertinence of the proposed model is shown not only on the dispersion
curves but also on the polar plots of the phase velocity which are
in good agreement with the analogous results obtained by means of
the discrete approach. The few differences that can be found between
the discrete and the continuum model are limited to few modes and
to high wavenumbers (small wavelengths). Such differences will be
treated in future works in which the role of the non-local inertia
terms will be investigated together with the case of non-vanishing
characteristic length $L_{c}>0$.

Future investigations will also focus on the mechanical characterization
of a larger class of band-gap metamaterials with the final aim of
the FE-implementation of morphologically complex band-gap metastructures.

Finally, the application of the relaxed micromorphic model to more
complex metamaterials including, e.g., piezoelectric effects will also
be investigated.

\section*{\label{subsec:C-macro 2}Appendix: equivalent dynamical determination of the
	macroscopic stiffness $\protect\cM$ }

In this appendix we show how the macroscopic parameters previously
obtained by simple static arguments can be equivalently computed using
the slopes of the acoustic dispersion curves close to the origin.
This equivalent method is very useful to make a strong connection
between the relaxed micromorphic model and classical elasticity, but
it does not add any extra feature to the fitting procedure presented
above. 

\subsection*{\label{subsec:Tangents-in-zero}Tangents in zero to the acoustic
	branches }

Let us consider the ``macroscopic'' Cauchy partial differential
system of equations 
\begin{equation}
\rho\,u_{,tt}-\textrm{Div}\left(\cM\,\sym\,\nabla\u\right)=0\label{modello macro}
\end{equation}
which is the limiting case of the relaxed micromorphic model when $L_c$ tends to zero. In this
classical case, it is possible to obtain an analytical expression
for the dispersion curves. In order to achieve this goal, we enter a generic plane wave function $\widetilde{u}\, e^{\gi\left(\left\langle \boldsymbol{k},x\right\rangle -\omega\,t\right)}$ in \eqref{modello macro} obtaining
\begin{align}\label{eq:space-der2}
\textrm{Div}\left[\cM\,\sym\,\nabla(\widetilde{u}\, e^{\gi\left(\left\langle \boldsymbol{k},x\right\rangle -\omega\,t\right)})\right] & =\left[(\cM)_{ijmn}\,\frac{1}{2}\left(\widetilde{u}_{m}k_{n}+\widetilde{u}_{n}k_{m}\right)\,\gi\,e^{\gi\left(\left\langle \boldsymbol{k},x\right\rangle -\omega\,t\right)}\right]_{,j}\nonumber \\
& =-\frac{1}{2}\left[(\cM)_{ijmn}\,\left(\widetilde{u}_{m}k_{n}+\widetilde{u}_{n}k_{m}\right)\,k_j\,e^{\gi\left(\left\langle \boldsymbol{k},x\right\rangle -\omega\,t\right)}\right]\nonumber \\
& =-\frac{1}{2}\left[\left(\cM\right)_{ijmn}k_{j}\widetilde{u}_{m}k_{n}+\left(\cM\right)_{ijmn}k_{j}\widetilde{u}_{n}k_{m}\right]e^{\gi\left(\left\langle \boldsymbol{k},x\right\rangle -\omega\,t\right)}\\
& =-\frac{1}{2}\left[\left(\cM\right)_{ijmn}k_{j}\widetilde{u}_{m}k_{n}+\left(\cM\right)_{ijnm}k_{j}\widetilde{u}_{n}k_{m}\right]e^{\gi\left(\left\langle \boldsymbol{k},x\right\rangle -\omega\,t\right)}\nonumber\\
&=-\left(\cM\right)_{ijmn}k_{j}\widetilde{u}_{m}k_{n}\,e^{\gi\left(\left\langle \boldsymbol{k},x\right\rangle -\omega\,t\right)},\nonumber 
\end{align}
for the static term and 
$$
\rho\,u_{,tt}=-\rho\,\omega^2\,\widetilde{u}\,e^{\gi\left(\left\langle \boldsymbol{k},x\right\rangle -\omega\,t\right)}=-\rho\,\omega^2\,\delta_{im}\,\widetilde{u}_m e^{\gi\left(\left\langle \boldsymbol{k},x\right\rangle -\omega\,t\right)}
$$
for the kinetic term. Thus, finally combining the two expressions we have that $\widetilde{u}\, e^{\gi\left(\left\langle \boldsymbol{k},x\right\rangle -\omega\,t\right)}$ satisfies the system \eqref{modello macro} if and only if 
$$
-\rho\,\omega^2\,\delta_{im}\,\widetilde{u}_m e^{\gi\left(\left\langle \boldsymbol{k},x\right\rangle -\omega\,t\right)}+\left(\cM\right)_{ijmn}k_{j}\widetilde{u}_{m}k_{n}\,e^{\gi\left(\left\langle \boldsymbol{k},x\right\rangle -\omega\,t\right)}=0
$$
and since the function $e^{\gi\left(\left\langle \boldsymbol{k},x\right\rangle -\omega\,t\right)}$ is always non-zero this gives
\begin{equation}
\left(\rho\,\omega^{2}\,\delta_{im}-\left(\cM\right){}_{ijmn} k_{j} k_{n}\right)\widetilde{u}_{m}=0
\end{equation}
and writing the wave vector as $\boldsymbol{k}=k\,\widehat{\boldsymbol{k}}$ with $\widehat{\boldsymbol{k}}=\big(\widehat{k}_{1},\widehat{k}_{2},\widehat{k}_{3}\big)$ unitary and $k=\left\| \boldsymbol{k} \right\| $, we arrive to
\begin{equation}\label{eq:autovali problem}
\left(\rho\,\omega^{2}\,\delta_{im}-k^{2}\,\left(\cM\right){}_{ijmn} \widehat{k}_{j} \widehat{k}_{n}\right)\widetilde{u}_{m}=0.
\end{equation}
The equation (\ref{eq:autovali problem}) is an eigenvalues problem
for the linear application $k^{2}\,\left(\cM\right){}_{ijmn}\widehat{k}_{j}\widehat{k}_{n}$.
The stated problem admits non-trivial solutions if and only if the
determinant of $\rho\,\omega^{2}\,\delta_{im}-k^{2}\,\left(\cM\right){}_{ijmn}\widehat{k}_{j}\widehat{k}_{n}$
is zero. In this way, we are interested in looking for couples $\left(k,\omega\right)$
such that 
\begin{equation}
\det\left(\rho\,\omega^{2}\,\delta_{im}-k^{2}\,\left(\cM\right){}_{ijmn}\widehat{k}_{j}\widehat{k}_{n}\right)=0.\label{eq:det0}
\end{equation}
Moreover, we are interested in studying this problem as a function
of the direction of propagation $\widehat{\boldsymbol{k}}$ of the
wave. In order to do this, it is convenient to introduce spherical
coordinates for the wave vector $\widehat{\boldsymbol{k}}\in\mathbb{S}^{2}$
(the unit sphere in $\R^{3}$): 
\begin{equation}
k_{1}=\sin\varphi\,\cos\vartheta,\qquad\qquad k_{2}=\sin\varphi\,\sin\vartheta,\qquad\qquad k_{3}=\cos\varphi,\label{eq:k}
\end{equation}
where $\vartheta\in\left[0,2\pi\right)$ is the polar angle and $\varphi\in\left[0,\pi\right]$
is the azimuthal angle. For the problem in the $\left(x_{1},x_{2},0\right)$
plane, the angle $\varphi$ is $\pi/2$, so
\begin{equation}
k_{1}=\cos\vartheta,\qquad\qquad k_{2}=\sin\vartheta,\qquad\qquad k_{3}=0,\label{eq:k-1}
\end{equation}
and
\[
\widehat{\boldsymbol{k}}\otimes\widehat{\boldsymbol{k}}=\begin{pmatrix}\cos^{2}\vartheta & \cos\vartheta\,\sin\vartheta & 0\\
\cos\vartheta\,\sin\vartheta & \sin^{2}\vartheta & 0\\
0 & 0 & 0
\end{pmatrix}.
\]
Let us now consider the Voigt representation of the tensor $\mathbb{C}_{\textrm{macro}}$
in the case of the tetragonal symmetry 
\[
\widetilde{\mathbb{C}}_{\textrm{macro}}=\begin{pmatrix}2\mu_{\textrm{macro}}+\lambda_{\textrm{macro}} & \lambda_{\textrm{macro}} & \lambda_{\textrm{macro}}^{*} & 0 & 0 & 0\\
\lambda_{\textrm{macro}} & 2\mu_{\textrm{macro}}+\lambda_{\textrm{macro}} & \lambda_{\textrm{macro}}^{*} & 0 & 0 & 0\\
\lambda_{\textrm{macro}}^{*} & \lambda_{\textrm{macro}}^{*} & (\widetilde{\mathbb{C}}_{\textrm{macro}})_{33} & 0 & 0 & 0\\
0 & 0 & 0 & (\widetilde{\mathbb{C}}_{\textrm{macro}})_{44} & 0 & 0\\
0 & 0 & 0 & 0 & (\widetilde{\mathbb{C}}_{\textrm{macro}})_{44} & 0\\
0 & 0 & 0 & 0 & 0 & \mu_{\textrm{macro}}^{*}
\end{pmatrix}.
\]
A direct calculation gives
\[
\left(\cM\right){}_{ijmn}\widehat{k}_{j}\widehat{k}_{n}=\left(\begin{array}{ccc}
\mu_{\textrm{macro}}^{\ast}\,\sin^{2}\vartheta+\cos^{2}\vartheta\left(\lambda_{\textrm{macro}}+2\mu_{\textrm{macro}}\right) & \cos\vartheta\sin\vartheta\left(\mu_{\textrm{macro}}^{\ast}+\lambda_{\textrm{macro}}\right) & 0\\
\cos\vartheta\sin\vartheta\left(\mu_{\textrm{macro}}^{\ast}+\lambda_{\textrm{macro}}\right) & \cos^{2}\vartheta\mu_{\textrm{macro}}^{\ast}+\sin^{2}\vartheta\left(\lambda_{\textrm{macro}}+2\mu_{\textrm{macro}}\right) & 0\\
0 & 0 & \big(\widetilde{\mathbb{C}}_{\textrm{macro}}\big)_{44}
\end{array}\right),
\]
thus, for $\rho\,\omega^{2}\,\delta_{im}-k^{2}\,\left(\cM\right){}_{ijmn}\widehat{k}_{j}\widehat{k}_{n}$,
we find
\[
\!\!\!\!\!\!\!\!\!\!\!\!\!\!\!\!\!\!\!\!\!\!\!\!\left(\begin{array}{ccc}
\rho\omega^{2}-k^{2}\left(\mu_{\textrm{macro}}^{\ast}\sin^{2}\vartheta+\cos^{2}\vartheta\left(\lambda_{\textrm{macro}}+2\mu_{\textrm{macro}}\right)\right) & -k^{2}\cos\vartheta\sin\vartheta\left(\mu_{\textrm{macro}}^{\ast}+\lambda_{\textrm{macro}}\right) & 0\\
-k^{2}\cos\vartheta\sin\vartheta\left(\mu_{\textrm{macro}}^{\ast}+\lambda_{\textrm{macro}}\right) & \rho\omega^{2}-k^{2}\left(\mum^{\ast}\cos^{2}\vartheta+\sin^{2}\vartheta\left(\lambda_{\textrm{macro}}+2\mu_{\textrm{macro}}\right)\right) & 0\\
0 & 0 & \rho\omega^{2}-\big(\widetilde{\mathbb{C}}_{\textrm{macro}}\big)_{44}k^{2}
\end{array}\right).
\]
In this way we can compute
\begin{align}
\det\left(\rho\,\omega^{2}\,\delta_{im}-k^{2}\,\left(\cM\right){}_{ijmn}\widehat{k}_{j}\widehat{k}_{n}\right)=\qquad\qquad\qquad\qquad\qquad\qquad\qquad\qquad\qquad\qquad\qquad\qquad\qquad\qquad\nonumber \\
=\left(\rho\,\omega^{2}-\big(\widetilde{\mathbb{C}}_{\textrm{macro}}\big)_{44}k^{2}\right)\Big[\left(\rho\,\omega^{2}-k^{2}\left(\cos^{2}\vartheta\left(\lam+2\mum\right)+\mum^{\ast}\sin^{2}\vartheta\right)\right)\qquad\qquad\qquad\label{eq:det2}\\
\left(\rho\,\omega^{2}-k^{2}\left(\sin^{2}\vartheta\left(\lam+2\mum\right)+\mum^{\ast}\cos^{2}\vartheta\right)\right)-k^{4}\sin^{2}\vartheta\cos^{2}\vartheta\left(\lam+\mum^{\ast}\right){}^{2}\Big].\nonumber 
\end{align}
The dispersion curves for the classical limit Cauchy model are obtained
solving the equation (\ref{eq:det0}), or equivalently (\ref{eq:det2}),
with respect to $\omega^{2}$. We call $\left\{ \pm\,\omega_{\textrm{macro},i}\left(k,\vartheta\right)\right\} _{i=1}^{3}$
the dispersion curves for the Cauchy continuum obtained by solving
(\ref{eq:det0}) for the special tetragonal case given in (\ref{eq:det2}).
A direct calculation shows that the positive solutions $\left\{ \omega_{\textrm{macro},i}\left(k,\vartheta\right)\right\} _{i=1}^{3}$
of (\ref{eq:det0}) for the tetragonal case are three straight lines
in the $\left(\omega,k\right)$ plane with slopes (here, and only
here, we use the abbreviations $\mu_{M}=\mum,$ $\lambda_{M}=\lam$,
$\mu_{M}^{*}=\mum^{*}$):
\begin{align}
a_{\mathrm{LA}} & =\sqrt{\frac{\lambda_{M}+\mu_{M}^{\ast}+2\mu_{M}+\sqrt{2\cos(4\vartheta)\left(\lambda_{M}+\mu_{M}\right)\left(\mu_{M}-\mu_{M}^{\ast}\right)+2\lambda_{M}\mu_{M}+\lambda_{M}^{2}+\mu_{M}^{\ast2}-2\mu_{M}\mu_{M}^{\ast}+2\mu_{M}^{2}}}{2\rho}},\label{relazioni pendenze}\\
a_{\mathrm{TA}} & =\sqrt{\frac{\lambda_{M}+\mu_{M}^{\ast}+2\mu_{M}-\sqrt{2\cos(4\vartheta)\left(\lambda_{M}+\mu_{M}\right)\left(\mu_{M}-\mu_{M}^{\ast}\right)+2\lambda_{M}\mu_{M}+\lambda_{M}^{2}+\mu_{M}^{\ast2}-2\mu_{M}\mu_{M}^{\ast}+2\mu_{M}^{2}}}{2\rho}},\nonumber \\
a_{\mathrm{TA}3} & =\sqrt{\frac{\left(\widetilde{\mathbb{C}}_{M}\right)_{44}}{\rho}}.
\end{align}
Note the complete absence of the Cosserat couple modulus $\mu_{c}$
in the latter formulas.

\begin{figure}[h]
	\begin{centering}
		\begin{tabular}{cc}
			\includegraphics[scale=0.5]{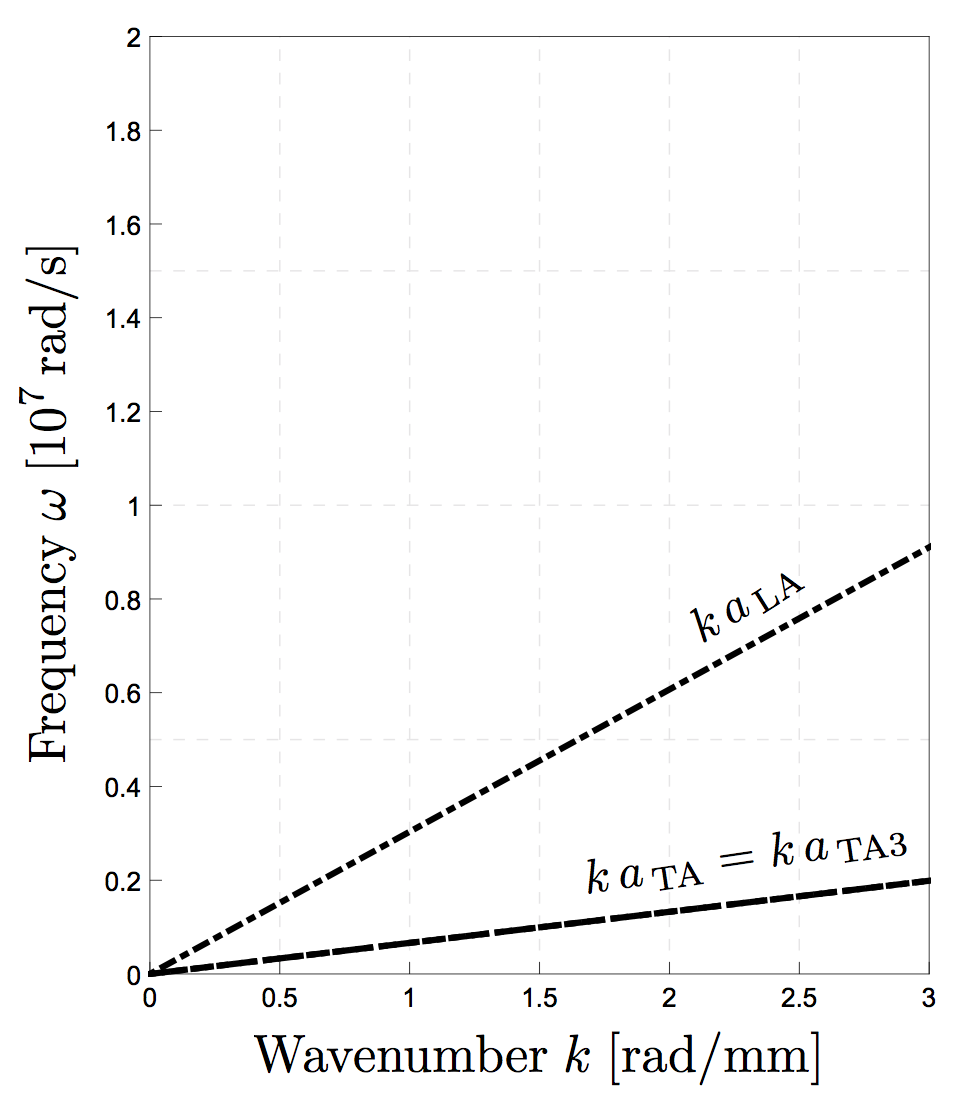} & \includegraphics[scale=0.5]{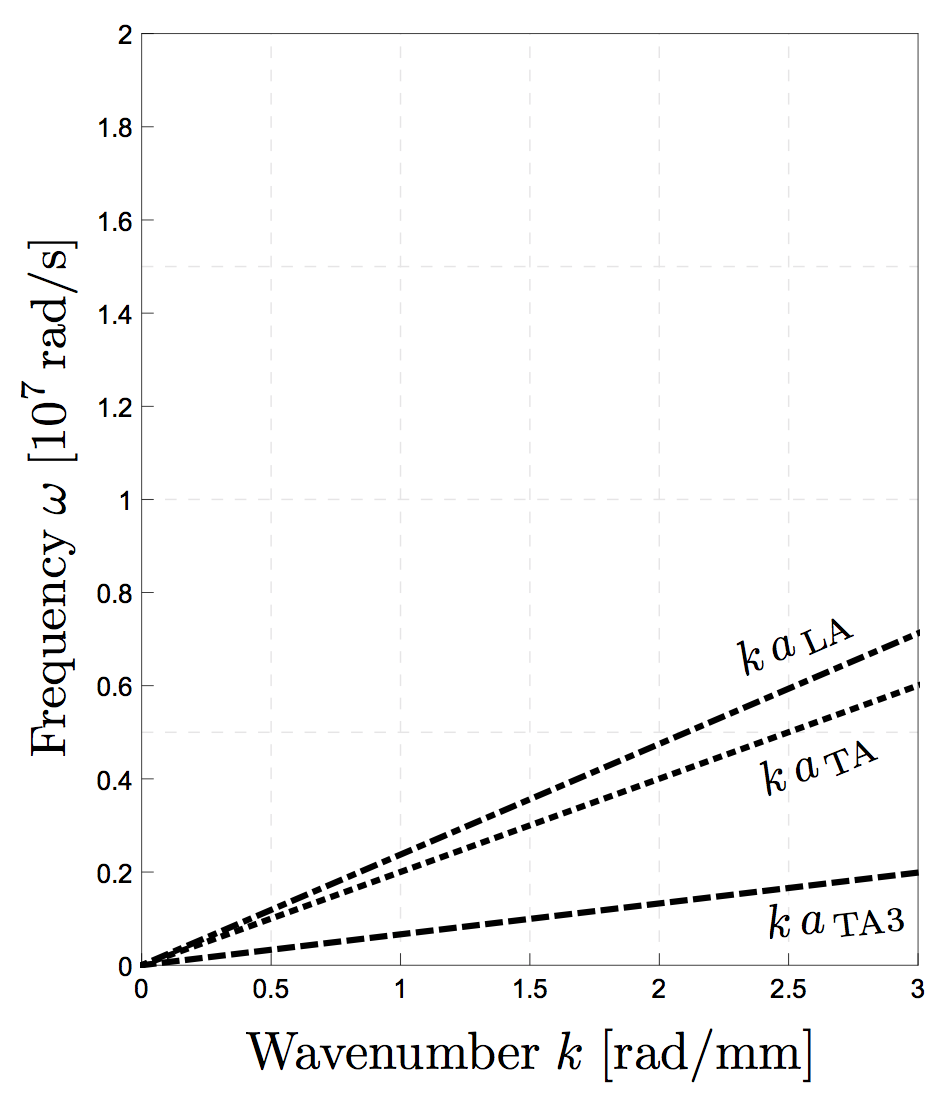}\tabularnewline
			(a) & (b)\tabularnewline
		\end{tabular}
		\par\end{centering}
	\caption{Dispersion branches for the limiting tetragonal Cauchy continuum with
		$\widehat{\boldsymbol{k}}=\left(1,0,0\right)$ (Fig.\,(a)) and $\widehat{\boldsymbol{k}}=\left(\sqrt{2}/2,\sqrt{2}/2,0\right)$
		(Fig.\,(b)).}
\end{figure}

Such dispersion curves are called (in-plane) longitudinal-acoustic
(LA), (in-plane) transverse-acoustic (TA) and (out-of-plane) transverse-acoustic (TA3). Since in this paper we are interested only in vibrations
in the $\left(x_{1},x_{2},0\right)$ plane, the third acoustic line
with slope $a_{\mathrm{TA}3}$ will not be considered for the fitting
procedure since it corresponds to out-of-plane vibrations.

One remarkable property of the relaxed micromorphic model is that
the slopes of its acoustic curves close to the origin, are exactly
given by the slopes of the acoustic lines (\ref{relazioni pendenze})
of the equivalent Cauchy continuum. More precisely, the slopes at
zero of the acoustic branches of the dispersion curves as obtained
via the relaxed micromorphic model can be computed by means of equations
(\ref{relazioni pendenze}) when using the identities (\ref{eq:mue-muh}).

\subsection*{Dynamical calculation of the macroscopic stiffness $\protect\cM$}

Based on the results of the previous Paragraph,  we
can compute the numerical values of the macroscopic parameters $\mum,\lam$
and $\mum^{*}$ which represent the measure of the macroscopic stiffness
of the considered tetragonal metamaterial. To this aim, considering
the two directions of propagation $\vartheta_{0}=0$ and $\vartheta_{1}=\pi/4$, and accounting for the following assumptions on the involved constitutive parameters
$$
\lam>0,\quad\mum>0,\quad\mum^{*}>0,\quad \lam+2\mum-\mum^{*}>0,
$$
we set up the following system of algebraic equations: 
\begin{align}
a_{\textrm{LA}}\left(\vartheta_{0},\lam,\mum,\mum^{*}\right)=\sqrt{\frac{2\mum+\lam}{\rho}} & =\overline{a}_{\textrm{LA}}\left(\vartheta_{0}\right),\nonumber \\
a_{\textrm{LA}}\left(\vartheta_{1},\lam,\mum,\mum^{*}\right)=\sqrt{\frac{\mum+\mum^{*}+\lam}{\rho}} & =\overline{a}_{\textrm{LA}}\left(\vartheta_{1}\right),\\
a_{\textrm{TA}}\left(\vartheta_{0},\lam,\mum,\mum^{*}\right)=\sqrt{\frac{\mu_{\textrm{macro}}^{*}}{\rho}} & =\overline{a}_{\textrm{TA}}\left(\vartheta_{0}\right),\nonumber \\
a_{\textrm{TA}}\left(\vartheta_{1},\lam,\mum,\mum^{*}\right)=\sqrt{\frac{\mu_{\textrm{macro}}}{\rho}} & =\overline{a}_{\textrm{TA}}\left(\vartheta_{1}\right).\nonumber 
\end{align}
This system of algebraic equations counts 4 equations and 3 unknowns.
We use the first 3 equations to calculate the unknowns $\lam,\mum,\mum^{*}$
and then plug the found values in the fourth equation. 

If our hypothesis according to which the metamaterial we are considering
has a tetragonal symmetry is correct, the fourth equation has to be
automatically satisfied. This is indeed the case.

The numerical values of the macroscopic parameters which are found
with the described procedure are given in Table \ref{Comsol parameters}.

\begin{table}[H]
	\begin{centering}
		\begin{tabular}{ccc}
			Bloch - Floquet &  & periodic homogenization\tabularnewline[2mm]
			\begin{tabular}{ccc}
				$\lam$ & $\mum$ & $\mum^{*}$\tabularnewline[1mm]
				\hline 
				\noalign{\vskip1mm}
				$\left[\mathrm{GPa}\right]$ & $\left[\mathrm{GPa}\right]$ & $\left[\mathrm{GPa}\right]$\tabularnewline[1mm]
				\hline 
				\hline 
				\noalign{\vskip1mm}
				$1.77$ & $5.95$ & $0.65$\tabularnewline[1mm]
			\end{tabular} &  & %
			\begin{tabular}{ccc}
				$\lam$ & $\mum$ & $\mum^{*}$\tabularnewline[1mm]
				\hline 
				\noalign{\vskip1mm}
				$\left[\textrm{GPa}\right]$ & $\left[\textrm{GPa}\right]$ & $\left[\textrm{GPa}\right]$\tabularnewline[1mm]
				\hline 
				\hline 
				\noalign{\vskip1mm}
				$1.74$ & $5.89$ & $0.62$\tabularnewline[1mm]
			\end{tabular}\tabularnewline
		\end{tabular}
		\par\end{centering}
	\caption{\label{Comsol parameters}Left: numerical values of the macroscopic
		parameters of the relaxed micromorphic model as obtained via the dynamical
		fitting and Bloch-Floquet analysis. Right: for comparison the values
		obtained by periodic homogenization.}
\end{table}

\section*{Acknowledgements}
Angela Madeo acknowledges funding from the French Research Agency ANR, "METASMART"(ANR-17CE08-0006) and the support from IDEXLYON in the framework of the
"Programme Investissement d?Avenir" ANR-16-IDEX-0005. All the authors acknowledge funding from the "Région
Auvergne-Rhône-Alpes" for the "SCUSI" project for international mobility France/Germany.
The work of I.D. Ghiba was supported by a grant of the \textquotedbl{}Alexandru
Ioan Cuza\textquotedbl{} University of Iasi, within the Research Grants
program, Grant UAIC, code GI-UAIC-2017-10. 
B. Eidel acknowledges support by the Deutsche Forschungsgemeinschaft (DFG) within the Heisenberg program (grant no. EI 453/2-1).

The authors thank the anonymous reviewers for their careful reading of our manuscript and their many insightful comments and suggestions. 

{\footnotesize{}\bibliographystyle{plain}
\bibliography{library-2}
}{\footnotesize \par}

\end{document}